\documentstyle[amsmath,amssymb,graphicx,rotate]{article}

\def\be{\begin{eqnarray}}
\def\ee{\end{eqnarray}}
\def\nn{\nonumber}

\def\Tr{{\rm Tr}\,}
\def\l[{\phantom.\!\![}
\def\hotR{{\mathcal R}} 

\hoffset= - 1.0in         

\title{{\bf Racah coefficients
and extended HOMFLY polynomials for all 5-, 6- and 7-strand braids}
\vspace{.5cm}}
\author{{\bf A.Anokhina}\footnote{ {\small {\it
MIPT, Dolgoprudny, Russia} and {\it ITEP, Moscow, Russia}};
anokhina@itep.ru}, \ {\bf A.Mironov}\footnote{ {\small {\it Lebedev
Physics Institute} and {\it ITEP, Moscow, Russia}}; mironov@itep.ru;
mironov@lpi.ru}, \ {\bf A.Morozov}\thanks{{\small {\it ITEP, Moscow,
Russia}}; morozov@itep.ru}, \ {\bf And.Morozov}\thanks{{\small {\it
Moscow State University} and {\it ITEP, Moscow, Russia}};
Andrey.Morozov@itep.ru}\date{ }}

\begin{document}

\setcounter{footnote}{3}

\setcounter{tocdepth}{3}

\maketitle

\vspace{-6.5cm}

\begin{center}
\hfill FIAN/TD-12/12\\
\hfill ITEP/TH-33/12
\end{center}

\vspace{5cm}

\begin{abstract}
Basing on evaluation of the Racah coefficients for $SU_q(3)$
(which supported the earlier conjecture of their universal form)
we derive explicit formulas for all the 5-, 6- and 7-strand Wilson
averages in the fundamental representation of arbitrary $SU(N)$
group (the HOMFLY polynomials). As an application, we list the
answers for all 5-strand knots with 9 crossings. In fact, the 7-strand
formulas are sufficient to reproduce all the HOMFLY polynomials from the katlas.org:
they are all described at once by a simple explicit formula with a very
transparent structure.
Moreover, would the formulas for the relevant $SU_q(3)$ Racah coefficients
remain true for all other quantum groups,
the paper provides a {\it complete}
description of the fundamental HOMFLY polynomials
for {\it all} braids with any number of strands.
\end{abstract}

\bigskip

\bigskip

\section{Introduction}

Knot theory now comes to the avant-scene of theoretical physics.
This was anticipated long ago \cite{Wit},
because it actually studies the Wilson loop averages in
a simplified (topological) version of Yang-Mills theory,
and is supposed to play the same crucially important role for
generic Yang-Mills studies
as topological strings play for the full string theory.
In the underlying $3d$ Chern-Simons theory \cite{Wit,CS}
many non-trivial effects are already present,
including sophisticated perturbation theory,
non-perturbative effects and various dualities:
just because of the topological nature they can be
sometime simplified and even eliminated by the gauge choices.
What is new today: a tremendous amount of "experimental"
data is available at \cite{katlas} about the Wilson loop
averages (called HOMFLY polynomials \cite{HOMFLY} within the Chern-Simons context).
Also effective theoretical methods are developed to calculate
these quantities based on studies of last years in adjacent
fields: group theory, matrix models, conformal theories
and AGT relations.
These advances led to a discovery of vast net of interrelations
between different HOMFLY polynomials including: various
difference equations \cite{Apol}, the AMM/EO topological recursion \cite{AMMEO,DFMGS}
and even some traces of integrability \cite{ammI}, which now need
to be systematized, understood and converted into more standard
forms common for other branches of science.
All this gives a new vim to the study of old,
but partly abandoned subjects like Racah coefficients
for quantum groups and character expansions.

Today, generic simple formulas are known for the HOMFLY polynomials for
all the 2,3,4-strand knots and links in the fundamental
representation \cite{ammII}; to write them down it turned
sufficient to know the representation theory of $SU_q(2)$, where the
Racah coefficients are well known \cite{RacahSL2} and were widely
used in other physical applications. In this letter
we report the extension of these results to the $5$-, $6$- and $7$-strand
cases, where $SU_q(3)$ representation theory (and its simple evident generalization)
is needed. Evaluation
{\it per se} of these Racah coefficients will be described in a
separate paper \cite{Ano}; what is important, this calculation
confirms the universal ansatz suggested in \cite{ammII}, which now
seems relatively safe to use for $m>7$ strands as well, this will
be done elsewhere. Note that {\bf the single $6$-strand formula
(\ref{6str})}, explicitly written in the present paper, {\bf is
sufficient to describe {\it all} the content of the
Rolfsen tables \cite{katlas}} concerning the HOMFLY polynomials in the
fundamental representation. Moreover, it is sufficient to
describe the HOMFLY polynomials of almost all knots in \cite{katlas}, since
all the tables are restricted to no more than 12
crossings, and all such knots, except just 17, admit the 6-strand
braid representation at most. The remaining 17 knots\footnote{
In accordance with \cite{katlas}, these are:
$12a_{0125}$; $12a_{0128}$; $12a_{0181}$; $12a_{0183}$; $12a_{0197}$; $12a_{0448}$; $12a_{0471}$; $12a_{0477}$;
$12a_0482$; $12a_{0690}$; $12a_{0691}$; $12a_{0803}$; $12a_{1124}$; $12a_{1127}$; $12a_{1166}$; $12a_{1202}$; $12a_{1287}$.
} require the $7$-strand
calculation which was also done with an evident generalization of the
Racah coefficients to the $SU_q(4)$ Young diagram [4,1,1,1] in (\ref{16}) and (\ref{17}).

\bigskip

According to \cite{ammI}, the HOMFLY polynomials (Wilson averages)
acquire the simplest form, if calculated in the gauge $A_0=0$
\cite{MS}\footnote{Our calculus is based on the approach by \cite{RT},
though that of \cite{inds} is, by essence, also very close.
} and expanded in the Schur functions $S_Q\{p\,\}$
(characters of the linear groups):
\be
{\cal H}_R^{\cal B}\{q|p\,\} = \sum_{Q\vdash m|R|} C_{RQ}^{\cal B}(q) S_Q\{p\,\}
\label{extHOMFLY}
\ee
where the expansion coefficients
\be
C_{RQ}^{\cal B}(q) =
\Tr_{N_{RQ}\times N_{RQ}} \left(\prod_{i=1}^\infty \prod_{j=1}^{m-1}
\hat{\cal R}_j^{a_{ij}}\right)
\label{coefHOMFLY}
\ee
Here\\

$\bullet$
${\cal B}$ is an $m$-strand braid
parameterized by a sequence of integers
\be
{\cal B} =
(a_{11},a_{12},\ldots,a_{1,m-1}|\,a_{21},a_{22},\ldots,a_{2,m-1}|\,a_{31},a_{32}\ldots)
\label{brapa}
\ee
These integers enter eq.(\ref{coefHOMFLY})
as powers of $\hat{\cal R}_j$-matrices along the braid: moving along the braid one first meets $a_{1k}$
links of $k$-th and $k+1$-th strands with some $k$, then $a_{1l}$ links of $l$-th and $l+1$-th strands with some $l\ne k$ (the sign of link is also taken into account). When one meets the links of the $k$-th and $k+1$-th strands for the second time, one associates with them the number $a_{2k}$.
The meaning of these integers can be understood from the figure
for the $3$-strand braid:

\vspace{0.7cm}

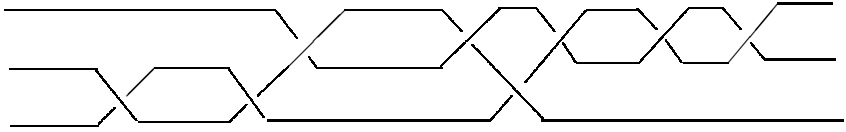
\unitlength 2.5pt
\linethickness{0.4pt}
\ifx\plotpoint\undefined\newsavebox{\plotpoint}\fi 
\begin{picture}(145.5,53)(0,0)
\put(19.5,34.5){\line(1,0){13.25}}
\put(41.25,43.25){\line(1,0){11.25}}
\put(19.25,43){\line(1,0){13.25}}
\put(38.75,35){\line(1,0){13.75}}
\put(61.25,43.25){\line(1,1){8.75}}
\put(70,52){\line(1,0){14.75}}
\put(18.5,52){\line(1,0){41}}
\multiput(59.5,52)(.033505155,-.043814433){97}{\line(0,-1){.043814433}}
\put(58.25,35.25){\line(1,0){33.75}}
\multiput(92,35.25)(.033505155,.038659794){97}{\line(0,1){.038659794}}
\multiput(64.5,45)(.03289474,-.04605263){38}{\line(0,-1){.04605263}}
\put(65.75,43.25){\line(1,0){19}}
\multiput(84.5,43.5)(.0346153846,.0336538462){260}{\line(1,0){.0346153846}}
\multiput(84.75,52)(.03370787,-.03651685){89}{\line(0,-1){.03651685}}
\multiput(52.5,43)(.033653846,-.046474359){156}{\line(0,-1){.046474359}}
\multiput(52.5,35)(.03353659,.03353659){82}{\line(0,1){.03353659}}
\multiput(56.75,39)(.035447761,.03358209){134}{\line(1,0){.035447761}}
\multiput(32.25,43)(.033602151,-.041666667){186}{\line(0,-1){.041666667}}
\multiput(32.75,34.75)(.03333333,.03333333){75}{\line(0,1){.03333333}}
\put(37,39){\line(1,1){4.25}}
\put(99.75,35.25){\line(1,0){45.75}}
\multiput(100,35.5)(-.0336990596,.0352664577){319}{\line(0,1){.0352664577}}
\multiput(97.25,41)(.0336363636,.04){275}{\line(0,1){.04}}
\put(106.5,52){\line(1,0){7.75}}
\put(121.25,44){\line(1,0){6.75}}
\put(128,44){\line(5,6){7.5}}
\put(135.5,53){\line(1,0){8.25}}
\put(93.25,52.25){\line(1,0){5.75}}
\multiput(99,52.25)(.03353659,-.04268293){82}{\line(0,-1){.04268293}}
\multiput(103,47)(.03333333,-.05){60}{\line(0,-1){.05}}
\put(105,44){\line(0,1){0}}
\put(105,44){\line(1,0){9.5}}
\multiput(114.5,44)(.033632287,.036995516){223}{\line(0,1){.036995516}}
\put(122,52.25){\line(1,0){5.25}}
\multiput(127.25,52.25)(.03353659,-.03963415){82}{\line(0,-1){.03963415}}
\multiput(131.5,47)(.03333333,-.04166667){60}{\line(0,-1){.04166667}}
\put(133.5,44.5){\line(1,0){10.75}}
\multiput(114.25,52.25)(.03370787,-.03651685){89}{\line(0,-1){.03651685}}
\multiput(121,44)(-.03333333,.04666667){75}{\line(0,1){.04666667}}
\end{picture}
\unitlength=1pt

\vspace{-2.2cm}

In this figure when moving from the left, one first meets 2 links of the second and the third strands, which gives (with account of sign) $a_{12}=-2$. Since the first and the second links do not cross at the beginning, one
puts $a_{11}=0$. Then, there are two links of the first and the second strands with
the opposite sign: $a_{21}=2$. The next link of the second and the third strands gives us $a_{22}=-1$,
and finally there are 3 links of the first and the second strands again, i.e. $a_{31}=3$.
This is knot $8_{10}$.

$\bullet$
$R$ is the representation parameterized by the Young diagram
$R = [r_1\geq r_2 \geq\ldots 0]$ with $|R|=\sum_k r_k$ boxes.
For the fundamental representation $R=\Box=[1]$.
The product $R^{\otimes m}$ is expanded in irreducible representations $Q$
of the size $|Q|=m|R|$:
\be
R^{\otimes m} = \sum_{Q\vdash m|R|} {\cal M}_{RQ}\otimes Q
\label{Rmexpan}
\ee
where ${\cal M}_{RQ}$ is the space of intertwining operators
$\ R^{\otimes m}\longrightarrow Q\ $ of dimension
$\ {\rm dim}\left({\cal M}_{RQ}\right) = N_{RQ}$.

$\bullet$ Finally, $\hat {\cal R}_j$ are quantum ${\cal
R}$-matrices. Originally
\be {\cal R}_j =
\underbrace{I\otimes\ldots\otimes I}_{j-1}\otimes{\cal R}
\otimes\underbrace{I\otimes\ldots\otimes I}_{m-j-1}
\ee
acts on
$R^{\otimes m}$, and is associated with the crossings of strands
$j$ and $j+1$. In the expansion (\ref{Rmexpan}), i.e. on the irreducible representations
its action is proportional to unity on
each $Q$, and, hence, reduces to an $N_{RQ}\times N_{RQ}$ matrix
acting on ${\cal M}_{RQ}$, which we denote through $\hat{\cal
R}_j$ (omitting indices $RQ$ which it actually depends on). These
matrices with different $j$ are related by orthogonal "mixing
matrices" $\hat {\cal U}_j$,
\be
\hat{\cal R}_j = \hat {\cal U}_j \hat{\cal R}_1
\hat {\cal U}^{-1}_j \label{mima}
\ee
made from the Racah coefficients of
$SU_q(\infty)$. Reduction to the Racah coefficients provides a natural
decomposition of $\hat {\cal U}_j$ into a product involving $m-2$ matrices (called
$\hat U,\hat V,\hat W,\hat Y,\hat Z$ in the present text, where $m=5,6,7$). In fact,
for given $m$ and $R$, in the product $R^{\otimes m}$ there is a diagram, at most, with
$ml(R)$ lines (where $l(R)$ denotes the number of lines in the Young diagram $R$), thus,
the representation theory of
$SU_q\left(ml(R)\right)$ is sufficient. However, one can use
the "mirror" symmetry under simultaneous changing $q\longrightarrow - \frac{1}{q} $
and transposing the Young diagram $Q\rightarrow \tilde
Q$ in order to reduce even this group. For instance, for $R=[1]=\Box$ one suffices to
consider $SU_q\left(\left[{m\over 2}\right]\right)$, where $[\ldots]$ denotes the integer part.
In particular, $SU_q(2)$ is enough for $R=\Box$ and $m=2,3,4$.
For $m=5,6$ one needs $SU_q(3)$, where the necessary Racah coefficients were
recently evaluated in \cite{Ano}. For $m=7$ one needs to know the Racah coefficients for $SU_q(4)$, however,
for a simple hook diagram $[4,1,1,1]$, when they are immediate, (\ref{16}).

\bigskip

Eq.(\ref{extHOMFLY}) defines the {\it extended} HOMFLY polynomials
\cite{ammI}, which depend on infinitely many time variables
$\{p_k\}$. This allows one to consider them as a kind of a
knot theory $\tau$-functions, though they belong to the ordinary
(free-fermion) KP/Toda family only for the torus knots.
Moreover, the extended polynomials depend on the braid
representation of a knot, and the knot invariants (conventional
HOMFLY polynomials) arise when the time variables are restricted
to the 1-dimensional {\it topological locus}
\be
p_k = p_k^* \equiv \frac{A^k-A^{-k}}{q^k-q^{-k}}\label{tolo}
\ee
where the Schur functions reduce to the quantum dimensions:
\be\label{Schurtolo}
S_Q\{p^*\} = \prod_{(k,l)\in Q} \frac{\left\{Aq^{k-l}\right\}}{\left\{q^{h_{k,l}}\right\}}
\ee
$h_{k,l}$ being a hook length and $\{x\} \equiv x-x^{-1}$.
Then, the HOMFLY polynomial
\be
H_R^{\cal K}(A|q) = \left(q^{4\varkappa_R}A^{|R|}\right)
^{-\sum_{ij} a_{ij}}
{\cal H}_R^{\cal B}\{p^*\}\label{framing}
\ee
with
the cut-and-join-operator eigenvalue $\varkappa_R = \sum_k(k-1)r_k$,
does not depend on the braid representation ${\cal B}$ of the knot\!/link ${\cal K}$.
In order to obtain the Wilson loop average for the gauge group
$SU(N)$ one should further put $A=q^N$.

\section{Racah coefficients for $[1]\times S\times[1] \rightarrow Q$}

Associativity of the tensor product:
\be
(R_1\times R_2)\times R_3 = R_1\times (R_2\times R_3) =
\sum_Q {\cal M}_{R_1R_2R_3}^Q \times Q
\ee
implies that there are two linear dependent bases in
${\cal M}_{R_1R_2R_3}^Q$:
\be
(R_1\times R_2)\times R_3 = \left(\sum_T {\cal M}_{R_1R_2}^T\times T\right)\times R_3
=  \left(\sum_T{\cal M}_{R_1R_2}^T\times{\cal M}_{TR_3}^Q\right)\times Q
\ee
\be
R_1\times (R_2\times R_3) = R_1\times \left(\sum_{T'} {\cal M}_{R_2R_3}^{T'}\times T'\right)
=  \left(\sum_{T'}{\cal M}_{R_1T'}^Q\times{\cal M}_{R_2R_3}^{T'}\right)\times Q
\ee
The two bases are related by the Racah matrix
\be
U_{R_1R_2R_3Q}^{TT'}
\ee
When $R_1=R_3=[1]$, {\bf this matrix is at most $2\times 2$},
since the two boxes can be added to $R_2$ to form a given $Q$
in at most two different ways (differing by permutation).

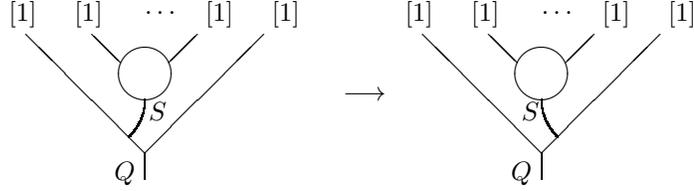
\unitlength 0.5pt
\begin{picture}(300,180)(-350,-40)
\put(0,0){\line(1,1){90}}
\put(0,0){\line(-1,1){90}}
\put(0,0){\line(0,-1){20}}
\put(-40,90){\line(1,-1){21.7}}
\put(40,90){\line(-1,-1){21.7}}
\qbezier(0,40)(0,25)(-12,12)
\put(0,60){\circle{38}}
\put(-103,100){\mbox{$[1]$}}
\put(-53,100){\mbox{$[1]$}}
\put(0,105){\mbox{$\ldots$}}
\put(46,100){\mbox{$[1]$}}
\put(96,100){\mbox{$[1]$}}
\put(3,25){\mbox{$S$}}
\put(-23,-20){\mbox{$Q$}}
\put(150,40){\mbox{$\longrightarrow$}}
\put(300,0){\line(1,1){90}}
\put(300,0){\line(-1,1){90}}
\put(300,0){\line(0,-1){20}}
\put(260,90){\line(1,-1){21.7}}
\put(340,90){\line(-1,-1){21.7}}
\qbezier(300,40)(300,25)(312,12)
\put(300,60){\circle{38}}
\put(197,100){\mbox{$[1]$}}
\put(247,100){\mbox{$[1]$}}
\put(300,105){\mbox{$\ldots$}}
\put(346,100){\mbox{$[1]$}}
\put(396,100){\mbox{$[1]$}}
\put(285,25){\mbox{$S$}}
\put(277,-20){\mbox{$Q$}}
\end{picture}
\unitlength 1pt

As suggested in \cite{ammII} and confirmed in \cite{Ano}, these
orthogonal matrices have the form (we change the notation
$R_2\rightarrow S$ to simplify the formulas)
\be
U_{[1]S[1]Q} =
\left(\begin{array}{cc} -u_{SQ} & \varepsilon_S\sqrt{\ 1-u_{SQ}^2}
\\ \sqrt{\ 1-u_{SQ}^2} & \varepsilon_S u_{SQ}
\end{array}\right)
\ee
where $\varepsilon_S = -1$ for $S=[1]$ and $\varepsilon_S = +1$ for all other $S$,
while $u_{SQ}^{-1}$ is a $q$-integer:
\be
u_{SQ} = \frac{1}{\left[k_{SQ}\right]}, \ \ \ \ \
\sqrt{\ 1-u_{SQ}^2} =
\frac{\sqrt{\left[k_{SQ}-1\right]\left[k_{SQ}+1\right]}}
{\left[k_{SQ}\right]}
\ee
where $[x]= \frac{q^x-q^{-x}}{q-q^{-1}} = \frac{\left\{q^x\right\}}{\{q\}}$
(the square brackets are also used to define the Young diagrams,
but this should not cause confusion because they appear in
different contexts). Remarkably, $[k]^2-1=[k+1][k-1]$.

According to \cite{ammII,Ano}, the actual values of $k_{SQ}$ are:

\bigskip

\noindent
\underline{$Q$-doublets descending from $S$:}

\be\label{16}
\begin{array}{c|c|cc}
S & Q && k_{SQ} \\
\hline\hline
\l[1] & [21] && 2 \\
\hline\hline
\l[2] & [31] && 3 \\
\hline
\l[11] & [211] && 3 \\
\hline\hline
\l[3] & [41] && 4 \\
\hline
\l[21]&[32]&&2\\
&[221]&&2\\
&[311]&&4 \\
\hline\hline
\l[4]&[51]&&5\\
\hline
\l[31]&[42]&&3\\
&[321]&&2\\
&[411]&&5\\
\hline
\l[22]&[321]&&4\\
\hline\hline
\end{array}
\ \ \ \ \ \ \ \ \ \ \ \ \ \ \ \
\begin{array}{c|c|cc}
S & Q && k_{SQ} \\
\hline\hline
\l[5]&[61]&&6\\
\hline
\l[41]&[52]&&4\\
&[421]&&2\\
&[511]&&6\\
\hline
\l[32]&[43]&&2\\
&[421]&&5\\
&[331]&&3\\
\hline
\l[311]&[421]&&3\\
&[4111]&&\boxed{6}\\
\hline\hline
\ldots&\ldots&\ldots
\end{array}
\ee

The general $SU_q(3)$-formula looks like
\be\label{Rc}
\boxed{
k_{SQ}=s_i-s_j+j-i}
\ee
where $i<j$ are the numbers of lines in the Young diagram
where the boxes are added
and $s_i>s_j$ are the lengthes of these lines. In particular, if one starts from a 1-hook diagram with $s_1=|S|-l+1$ and
$s_2=s_3=\ldots=s_l=1$ and puts the new boxes to the ends obtaining a new 1-hook diagram, then the simple rule
$k_{SQ}=|S|-l+1+l+1-1=|S|+1=|Q|-1$ works.
Here
$|S|$ is the {\it depth} (or the level) of the Racah coefficient,
see the next section. This rule is immediately generalized to higher rank groups, for instance, the hook Young
diagram (the boxed number in the table) would require $SU_q(4)$.

Formula (\ref{Rc}) can be derived by "the brute force" following the way described in detail in \cite{ammII}. It is done
in \cite{Ano}. However, one can guess it from its values at $q=1$ and "check" by comparing the HOMFLY polynomials
for 6- and 7-strands (i.e. for groups $SU_q(3)$ and $SU_q(4)$) with known results.

The mixing matrices $\hat {\cal U}_j$ in (\ref{mima}) with $j=1,\ldots,m-1$,
are products of involving $j-1$ elementary Racah matrices,
which in their turn are made from such $2\times 2$ blocks
complemented by unit matrices with the sign factors $v_{SQ} = \pm 1$,
depending on $S$ and $Q$:

\bigskip

\noindent
\underline{$Q$-singlets descending from $S$:}
\be\label{17}
\begin{array}{l|c|cr}
S & Q && v_{SQ} \\
\hline\hline
[1] & [3] && 1 \\
& [111] && 1 \\
\hline\hline
[2] & [4] && 1 \\
& [22] && 1 \\
& [211] && -1\\
\hline
[11] & [31] && 1\\
&[22]&&-1\\
\hline\hline
[3] & [5] && 1\\
& [32] && 1 \\
&[311]&&-1\\
\hline
[21]&[41]&&1\\
\hline
[111]&[311]&& 1 \\
&[221]&&-1\\
\hline\hline
\end{array}\ \ \ \ \ \ \ \ \ \  \ \ \ \ \ \ \
\begin{array}{l|c|cr}
S & Q && v_{SQ} \\
\hline\hline
[4]&[6]&&1\\
&[42]&&1\\
&[411]&&-1\\
\hline
[31]&[51]&&1\\
&[33]&&1\\
\hline
[22]&[42]&&1\\
&[33]&&-1\\
&[222]&&1\\
\hline\hline
\end{array}\ \ \ \ \ \ \ \ \ \  \ \ \ \ \ \ \
\begin{array}{l|c|cr}
S & Q && v_{SQ} \\
\hline\hline
[5]&[7]&&1\\
&[52]&&1\\
&[511]&&-1\\
\hline
[41]&[61]&&1\\
&[43]&&1\\
\hline
[32]&[52]&&1\\
&[322]&&1\\
\hline
[311]&[511]&&1\\
&[331]&&1\\
&[322]&&-1\\
\hline\hline
\ldots&\ldots&\ldots
\end{array}
\ee
In general, the unit matrices correspond to the case when the order of adding the boxes can not be changed,
with $v_{SQ}=1$ when the boxes are added to the same line and $v_{SQ}=-1$ when the boxes are added to the
neighbor lines of equal lengthes. This fact does not depend on the rank of the gauge group.

\section{Explicit construction of mixing matrices}

In this section we describe a very simple and obvious procedure
to build up an arbitrary mixing matrix.
It is provided in the form of a product of elementary Racah matrices,
each made out of $1\times 1$ and $2\times 2$ blocks with
entries listed in the tables (\ref{16}) and (\ref{17}).
$SU_q(L)$ group theory is needed {\it only} to find the
entries of these tables.
Once this is done, the problem of finding the fundamental HOMFLY polynomials
will be {\bf solved completely}.
So far, our reliable knowledge is sufficient to describe all $m\leq 7$ braids.
Explicit formulas for the mixing matrices are given below in s.4 and the Appendices,
here we briefly describe the way to construct them.

\subsection{Decomposition of mixing matrices}

The mixing matrix $\hat{\cal U}_j$ converts the $\hat{\cal R}$ matrix,
acting on the first pair of $[1]$ in the product $[1]^{\times m}$
into that acting on the $j$-th pair:
\be
\underbrace{[1]\times[1]}_{\hat{\cal R}_1}\times [1]\times\ldots
\ \ \stackrel{\hat{\cal U}_j}{\longrightarrow}\ \
\overbrace{[1]\times[1]\times\ldots\times
\underbrace{[1]\times[1]}_{\hat{\cal R}_j}}^{j+1}
\times\ldots
\ee
$\hat{\cal R}_1$ intertwines the first two $[1]$'s in the product and does
not affect any other, i.e. in order to define $\hat{\cal R}_1$ it is important to
multiply at the beginning the first two $[1]$'s and after that all other can be
multiplied in an arbitrary order. For $\hat{\cal R}_j$ the first to be
multiplied is the $j$-th pair, and all other do not matter.

In other words, in order to convert $\hat{\cal R}_1$ into $\hat{\cal R}_2$
it is enough to consider the Racah transform
$$
(1\times 1)\times 1 \ \ \stackrel{U}{\longrightarrow}\ \ 1\times (1\times 1)
$$
(from now on, we omit the square brackets to simplify the formulas,
at least, a little).
Similarly, to convert $\hat{\cal R}_1$ into $\hat{\cal R}_3$ one needs the three steps:
$$
\begin{array}{ccccc}
(1\times 1)\times 1\times 1 &\cong &\Big((1\times 1)\times 1\Big)\times 1
&\ \ \stackrel{U}{\longrightarrow}\ \ &\Big(1\times (1\times 1)\Big)\times 1
\\ \\
&&&&\downarrow \ V
\\ \\
1\times 1\times (1\times 1) &\cong &1\times\Big(1\times(1\times 1) \Big)
&\ \ \stackrel{U}{\longleftarrow}\ \ &1\times \Big((1\times 1)\times 1\Big)
\end{array}
$$
The sign $\cong$ means that the two expressions are equivalent
from the point of view of the action of $\hat{\cal R}_1$ or $\hat{\cal R}_3$.

When we move the bracket $(1\times S)\times 1 \ \longrightarrow \ 1\times(S\times 1)$,
we say that we do this at the {\it depth} $|S|$ ($|S|$ denotes the number of boxes in the Young diagram $S$;
since here we obtain $S$ from products of representations $[1]$, it comes from $[1]^{\otimes |S|}$).
Following \cite{ammII}, the transition of depth $1$ is denoted by $U$,
that of the depth $2$ by $V$, and we use $W$, $Y$ and $Z$ for the depths
$3$, $4$ and $5$.
Thus, the mixing matrix $\hat{\cal U}_2$ is of the type $U$,
$\hat{\cal U}_3$ is the combination of three Racah matrices $UVU$,
$\hat{\cal U}_4$ will be $UVUWVU$ and so on.
In fact, there are combinations with different order of the Racah matrices for a $\hat{\cal U}_j$ with $j\geq 3$,
but the Racah matrices with depth difference exceeding one commute and,
thus, all the seemingly different representations are in fact the same.

\subsection{Representation tree}

The main object that we need in order to construct the Racah matrices $U,V,W,Y,Z,\ldots$
is the {\it representation tree}:
{\footnotesize
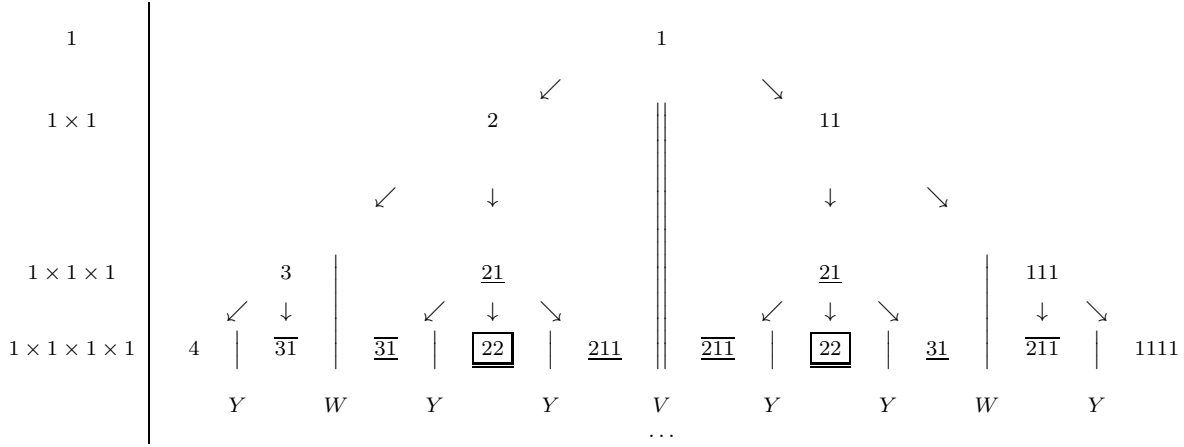
$$
\begin{array}{c|ccccccccccccccccccccccc}
&&\ &&&&&&&&& &&&&&&&\\
1&&&&&&&&&&&1 &&&&&&&\\
&&&&&&&&&&& &&&&&&&\\
&&&&&&&&&\swarrow&& &&\searrow&&&&&&\\
1\times 1&&&&&&&&2&&&\Big|\Big| &&&11&&&&&\\
&&&&&&&&&&&\Big|\Big| &&&&&&&&\\
&&&&&&\swarrow&&\downarrow&&&\Big|\Big| &&&\downarrow&&\searrow&&&\\
&&&&&&&&&&&\Big|\Big| &&&&&&&&\\
1\times 1\times 1&&&&3&\Big|&&&\underline{21}&&&\Big|\Big| &&&\underline{21}&&&\Big|&111&&\\
&&&\swarrow&\downarrow&\Big|&&\swarrow&\downarrow&\searrow&&\Big|\Big|
  &&\swarrow&\downarrow&\searrow&&\Big|&\downarrow&\searrow&\\
1\times 1\times 1\times 1
&&4&\Big|&\overline{31}&\Big|&\overline{\underline{31}}&\Big|
&\underline{\boxed{22}}&\Big|&\underline{211}&\Big|\Big|
  &\overline{\underline{211}}&\Big|&\underline{\boxed{22}}
  &\Big|&\underline{31}&\Big|&\overline{211}&\Big|& 1111 \\
&&&&&&&&&&&&& &&&&&&&\\
&&&Y&&W&&Y&&Y&&V &&Y&&Y&&W&&Y&\\
&&&&&&&&&&& \ldots
\end{array}
$$}
We actually need this tree at least up to the level $1^{\times 7}$,
but the space is not enough to present it here, in what follows we
draw also a fragment of the tree at the next levels.

\subsection{The structure of Racah matrices}

\subsubsection{Depth one ($U$)}

The first Racah matrix, $U$, appears at the level of $1\times 1\times 1$,
and it mixes the two underlined representations $[21]$.
This matrix is $2\times 2$ and explicitly given by (\ref{16}) with
$k_{[1],[21]} = 2$, this $2=depth +1$.
The {\it same} $2\times 2$ matrix will mix all the {\it descendants}
of these two $[21]$ at lower levels: the two underlined $[31]$,
the two underlined $[22]$ and the two underlined $[211]$.
There is still one other $[31]$ and one other $[211]$ at level $4$,
which are not underlined, and are not affected by the matrix $U$,
which has $v_{[1],[3]}= v_{[1],[111]}=1$ at the corresponding positions:
see eq.(\ref{23}) below.
The {\it same} $2\times 2$ blocks with the same $k_{[1],[21]}=2$
will appear in the $U$-matrices for all other descendants of the two $[21]$
at lower levels. If the same representations appear, which are the descendants
of $[3]$ and $[111]$, the corresponding entries of $U$-matrices are
$1\times 1$ and equal to $v_{[1],[3]}=1$ and $v_{[1],[111]}=1$.

To finalize the notational agreements, $U_{[31]}$ has three rows and three columns,
corresponding to the three appearances of $[31]$ in the representation tree,
and they are ordered just as in the tree: the first remains unaffected by $U$,
the second and the third are mixed.
For $[211]$ the $2\times 2$ block would involve the first two rows and columns,
but we do not actually need this mixing matrix, because the contribution
of $[211]$ to the HOMFLY polynomial is the mirror image of the $[31]$ contribution.

\subsubsection{Depth two ($V$)}

The second (depth-two) Racah matrix, $V$, describes the transition
$$
\Big([1]\times \underbrace{([2]+[11])}_S\Big) \times [1] \ \stackrel{V}{\longrightarrow}
\ [1]\times \Big(\underbrace{([2]+[11])}_S\times [1]\Big)
$$
This means that $[2]$ and $[11]$ are {\it not} affected by $V$,
and so are all their descendants: there is a "wall" separating
representations, which can be mixed by $V$: it is the vertical double line
in the picture, labeled by $V$.
This means that at the level $1^{\times 4}$ only the two overlined $[31]$
and the two overlined $[211]$ can be mixed,
this time in the $[31]$ sector the $2\times 2$ block involves the first
two rows and columns. And also the corresponding $k_{[2],[31]}=3$ is now different.
The same $V$ with the same $k_{[2],[31]}=3$ will mix all the descendants of
these two $[31]$ at all the lower levels. All other representations in
the left half of the tree will remain intact under $V$ (at most, change sign,
if the corresponding $v_{[2],Q}=-1$
(according to (\ref{17}), this happens only for $Q=[211]$).

\subsubsection{Depth three ($W$)}

The depth-three Racah matrix $W$ is associated with the transition
$$
\Big([1]\times \underbrace{([3]+2\cdot[21]+[111])}_S\Big) \times [1] \
\stackrel{W}{\longrightarrow}
\ [1]\times \Big(\underbrace{([3]+2\cdot[21]+[111])}_S\times [1]\Big)
$$
and there are now three $W$-"walls" separating representations, which
can {\it not} be mixed by $W$ (one of the $W$-walls coincides with the $V$-wall).
To see what can be mixed by $W$, we draw now the next levels, but only for the
left half of the representation tree:

{\footnotesize
\centerline{
$
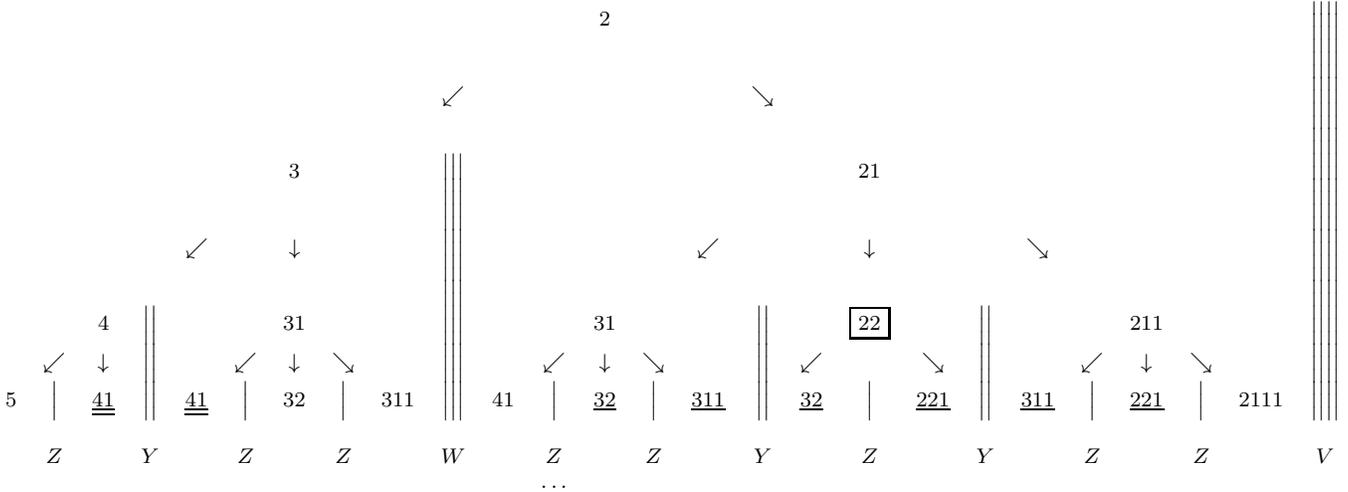
\begin{array}{ccccccccccccccccccccccccccccccccccccccccccccccc}
&&\ &&&&&&&&& &&&&&&&\\
&&&&&&&&&&&&2&&&&&&&&&&&&&\Big|\Big|\Big|\Big| \\
&&&&&&&&&&&&&&&&&&&&&&&&&\Big|\Big|\Big|\Big| \\
&&&&&&&&&\swarrow&&&&&&\searrow&&&&&&&&&&\Big|\Big|\Big|\Big|  \\
&&&&&&&&&&&&&&&&&&&&&&&&&\Big|\Big|\Big|\Big|  \\
&&&&&&3&&&\Big|\Big|\Big|&&&&&&&&21&&&&&&&&\Big|\Big|\Big|\Big|  \\
&&&&&&&&&\Big|\Big|\Big|&&&&&&&&&&&&&&&& \Big|\Big|\Big|\Big|\\
&&&&\swarrow&&\downarrow&&&\Big|\Big|\Big|
&&&&&\swarrow&&&\downarrow&&&\searrow&&&&&\Big|\Big|\Big|\Big|\\
&&&&&&&&&\Big|\Big|\Big|&&&&&&&&&&&&&&&& \Big|\Big|\Big|\Big|\\
&&4&\Big|\Big|&&&31&&&\Big|\Big|\Big|&&&31&&&\Big|\Big|
&&\boxed{22}&&\Big|\Big|&&&211&&&\Big|\Big|\Big|\Big|\\
&\swarrow&\downarrow&\Big|\Big|&&\swarrow&\downarrow&\searrow&&\Big|\Big|\Big|
&&\swarrow&\downarrow&\searrow&&\Big|\Big|&\swarrow&&\searrow&\Big|\Big|
&&\swarrow&\downarrow&\searrow&&\Big|\Big|\Big|\Big| \\
5 &\Big| & \underline{\underline{41}} &\Big|\Big|
&\underline{\underline{41}} &\Big|& 32 &\Big|& 311 &\Big|\Big|\Big|
& 41 &\Big|& \underline{32} &\Big|& \underline{311} &\Big|\Big|& \underline{32}
&\Big|& \underline{221} &\Big|\Big|
& \underline{311}&\Big|& \underline{221} &\Big|& 2111 &\Big|\Big|\Big|\Big|\\
&&&&&&&&&&&&&&&&&&&&&&&&& \\
&Z&&Y&&Z&&Z&&W&&Z&&Z&&Y&&Z&&Y&&Z&&Z&&V  \\
&&&&&&&&&&& \ldots
\end{array}
$
}}
Now it is clear that $W$-mixed at the fifth level are
the two double-underlined $[41]$ to the left of the $W$-wall,
and three underlined pairs to the right of it:
$[32]$, $[311]$ and $[221]$, and this will generate
exactly the same $W$-mixing
of all their descendants at lower levels.

The new thing is that the numbers $k_{SQ}$ at the same depth can now be different.
Deviation from the simple rule $k = depth+1$ occurs when mixed are descendants
of a non-hook diagram, namely $[22]$.
The general rule of this deviation is the only remaining uncertainty in description of the
mixing matrices, but in this particular case it is explicitly done in
the table (\ref{16}).
Note that the two hook diagrams $[311]$ are {\it not}
the descendants of $[22]$,
therefore, their mixing obeys the rule $k=|S|+1$.
The two non-hook diagrams $[32]$ and $[221]$ are $[22]$-descendants,
and for them the $k$-numbers are smaller.

Another new thing is that at level five the mixing occurs for the first time
between the descendants of $[21]$ only, namely $[32]$, $[311]$ and $[221]$.
Since there are two $[21]$ in the representation tree, this means that there are
two copies of these mixing pairs, and the Racah matrices contain two
identical $2\times 2$ blocks, not a single such block, as it happened
at levels three and four.
However, at level five these two blocks are still identical.

The last new phenomenon is the occurrence of $2\times 2$ blocks with
two or more different $k$-numbers within one and the same Racah matrix.
This happens when the same representation can appear both in the product
of two hook and of non-hook diagrams.
For the first time this happens at level $6$, where
one and the same representation, namely $[42]$ or $[321]$, appears
in two copies at different sides of the wall, and this wall can be either
$W$ or $Y$. Therefore, the four Racah matrices
$W_{[42]}$, $Y_{[42]}$, $W_{[411]}$, $Y_{[411]}$ can contain $2\times 2$
blocks with different $k$-numbers, and this is indeed the case, see Appendix 1.

\subsubsection{Depth four and five ($Y$ and $Z$)}

These can be analyzed in exactly the same way, but nothing essentially
new happens, just matrices become bigger and bigger.
In general, say, the hook representation $[k,1^{m-k}]$
appears $\frac{(m-1)!}{(k-1)!(m-k)!}$ times in the expansion of $[1]^{\times m}$,
i.e. for $m$ strands. The mixing matrices for symmetric hooks are the biggest,
but they are easily predictable: all the $k$-numbers for all hook diagrams
depend only on the depth: $k_{\rm hook} = |S|+1=|Q|-1$.
This fact allows us to define the mixing matrices for the
representation $[4111]$: the only one for $7$ strands,
which is not controlled by representation theory of $SU_q(3)$,
and for the "honest" calculation $SU_q(4)$ is needed.
It is needed anyway for the non-hook diagrams with four columns,
which contribute to calculations for $m\geq 8$.
To have the complete description of the mixing matrices
(of their Racah constituents, to be exact) one needs a generic formula
for {\it all} the elements of the tables (\ref{16}) and (\ref{17}).
It can easily happen that formula (\ref{Rc})
remains valid for $m\geq 8$, but this remains to be checked.
Knot theory itself does not help to make a check, because
non-hook diagrams do not contribute in the case of the torus knots
(in the fundamental representation), and not much is known independently
about non-torus knots for $m\geq 8$: as we already mentioned there
is none of this kind in \cite{katlas}.

\section{Explicit formula for extended HOMFLY polynomials}

We are now ready to provide explicit expressions for
the HOMFLY polynomials (\ref{extHOMFLY}): it remains to
substitute into (\ref{coefHOMFLY}) concrete expressions
for $\hat{\cal R}_j$ through $\hat{\cal R} = \hat{\cal R}_1$
and Racah matrices. We also use an operation
which changes $q\longrightarrow - \frac{1}{q} $
and transposes the Young diagram $Q \longrightarrow \tilde Q$.
This symmetry effectively used in \cite{DMMSS,IMMMfe,iammIII}
was named mirror in \cite{GS}. It
not only simplifies the formulas,
it allows one to restrict consideration to the Young diagrams with
no more than $m/2$ (rather than $m$) columns, for which the
Racah coefficients are provided by the theory of $SU_q(m/2)$
rather than $SU_q(m)$ group (of course $m/2 \rightarrow (m+1)/2$
for $m$ odd).

\subsection*{m=2}

\be
{\cal H}_{_\Box}^{(a)} = q^a S_{[2]} +\left(-\frac{1}{q}\right)^a S_{[11]} = q^a S_{[2]} + (mirror)
\ee

\subsection*{m=3}

\be
{\cal H}_{_\Box}^{(a_1b_1|a_2b_2|\ldots)} = q^{\sum_i (a_i+b_i)} S_{[3]}
+ (mirror)
+ 
\left( \Tr_{2\times 2}\prod_i \left(
\hat{\cal R}_{[21]}^{a_i}\hat U_{[21]} \hat{\cal R}_{[21]}^{b_i}\hat U_{[21]}^\dagger\right)
\right)\cdot S_{[21]}
\ee
In this particular case $\ (mirror) = \left(-\frac{1}{q}\right)^{\sum_i(a_i+b_i)} S_{[111]}\ $
and
\be
\hat{\cal R}_{[21]} = \left(\begin{array}{cc} q \\ & -\frac{1}{q} \end{array}\right),
\ \ \ \ \ \ \ \
\hat U_{[21]} = \left(\begin{array}{cc} -\frac{1}{[2]} & -\frac{\sqrt{[3]}}{[2]} \\ \\
\frac{\sqrt{[3]}}{[2]}& -\frac{1}{[2]} \end{array}\right)
\ee

\subsection*{m=4}

\be
{\cal H}_{_\Box}^{(a_1b_1c_1|a_2b_2c_2|\ldots)} = q^{\sum_i (a_i+b_i+c_i)} S_{[4]}
+ S_{[31]} \cdot \Tr_{3\times 3} \prod_i \left(
\hat{\cal R}_{[31]}^{a_i}\hat U_{[31]} \hat{\cal R}_{[31]}^{b_i}\hat V_{[31]}\hat U_{[31]}
\hat{\cal R}_{[31]}^{c_i}\hat U_{[31]}^\dagger\hat V_{[31]}^\dagger\hat U_{[31]}^\dagger\right)  + \nn \\
+ \Big(mirror\Big) \
+\  S_{[22]} \cdot \Tr_{2\times 2} \prod_i \left(
\hat{\cal R}_{[22]}^{a_i}\hat U_{[22]} \hat{\cal R}_{[22]}^{b_i}\hat V_{[22]}\hat U_{[22]}
\hat{\cal R}_{[22]}^{c_i}\hat U_{[22]}^\dagger\hat V_{[22]}^\dagger \hat U_{[22]}^\dagger\right)
\ee
Now $\ (mirror)\ $ contains contributions from two diagrams $Q=[211]$ and $Q=[1111]$,
while
\be\label{23}
\hat{\cal R}_{[31]} = \left(\begin{array}{ccc} q \\ & q \\ && -\frac{1}{q}\end{array}\right),
\ \ \ \
\hat U_{[31]} =
\left(\begin{array}{ccc} 1 \\ & -\frac{1}{[2]} & -\frac{\sqrt{[3]}}{[2]} \\ \\
& \frac{\sqrt{[3]}}{[2]}& -\frac{1}{[2]} \end{array}\right),
\ \ \ \
\hat V_{[31]} =
\left(\begin{array}{ccc}   -\frac{1}{[3]} & \frac{\sqrt{[2][4]}}{[3]} &  \\ \\
 \frac{\sqrt{[2][4]}}{[3]}& \frac{1}{[3]} & \\ \\ && 1 \end{array}\right)
\ee
and
\be
\hat{\cal R}_{[22]} = \left(\begin{array}{cc} q \\  & -\frac{1}{q}\end{array}\right),
\ \ \ \ \ \ \
\hat U_{[22]} =
\left(\begin{array}{cc}   -\frac{1}{[2]} & -\frac{\sqrt{[3]}}{[2]} \\ \\
  \frac{\sqrt{[3]}}{[2]}& -\frac{1}{[2]} \end{array}\right),
\ \ \ \ \ \ \
\hat V_{[22]} = \left(\begin{array}{cc}   1  \\
  & -1 \end{array}\right)
\ee

\subsection*{m=5}

$$
{\cal H}_{_\Box}^{(a_1b_1c_1d_1|a_2b_2c_2d_2|\ldots)} = q^{\sum_i (a_i+b_i+c_i+d_i)} S_{[5]}+
$$$$
+ S_{[41]} \cdot \Tr_{4\times 4} \prod_i \left(
\hat{\cal R}_{[41]}^{a_i}\hat U_{[41]} \hat{\cal R}_{[41]}^{b_i}\hat V_{[41]}\hat U_{[41]}
\hat{\cal R}_{[41]}^{c_i}\hat W_{[41]}\hat V_{[41]}\hat U_{[41]} \hat{\cal R}_{[41]}^{d_i}
\hat U_{[41]}^\dagger\hat V_{[41]}^\dagger\hat W_{[41]}^\dagger\hat U_{[41]}^\dagger\hat V_{[41]}^\dagger\hat U_{[41]}^\dagger\right)  + $$
$$
+ S_{[32]} \cdot \Tr_{5\times 5} \prod_i \left(
\hat{\cal R}_{[32]}^{a_i}\hat U_{[32]} \hat{\cal R}_{[32]}^{b_i}\hat V_{[32]}\hat U_{[32]}
\hat{\cal R}_{[32]}^{c_i}\hat W_{[32]}\hat V_{[32]}\hat U_{[32]} \hat{\cal R}_{[32]}^{d_i}
\hat U_{[32]}^\dagger\hat V_{[32]}^\dagger\hat W_{[32]}^\dagger\hat U_{[32]}^\dagger\hat V_{[32]}^\dagger\hat U_{[32]}^\dagger\right)
+\  \Big(mirror\Big) \ +
$$ \vspace{-0.2cm}
\be
+\  S_{[311]} \cdot \Tr_{6\times 6} \prod_i \left(
\hat{\cal R}_{[311]}^{a_i}\hat U_{[311]} \hat{\cal R}_{[311]}^{b_i}\hat V_{[311]}\hat U_{[311]}
\hat{\cal R}_{[311]}^{c_i}\hat W_{[311]}\hat V_{[311]}\hat U_{[311]} \hat{\cal R}_{[311]}^{d_i}
\hat U_{[311]}^\dagger\hat V_{[311]}^\dagger\hat W_{[311]}^\dagger\hat U_{[311]}^\dagger\hat V_{[311]}^\dagger\hat U_{[311]}^\dagger\right)
\label{5str}
\ee
This $\ (mirror)\ $ contains contributions from three diagrams $\ Q=[2111],\ [221]$ and $[11111]$,
while
\be
\hat{\cal R}_{[41]} = \left(\begin{array}{cccc} q \\ & q \\ && q\\ &&& -\frac{1}{q}\end{array}\right),
\ \ \ \
\hat U_{[41]} =
\left(\begin{array}{cccc} 1 \\ & 1 \\ & & -\frac{1}{[2]} & -\frac{\sqrt{[3]}}{[2]} \\ \\
& &  \frac{\sqrt{[3]}}{[2]}& -\frac{1}{[2]} \end{array}\right),
\ \ \ \ \nn \\
\hat V_{[41]} =
\left(\begin{array}{cccc}  1 \\ \\ & -\frac{1}{[3]} & \frac{\sqrt{[2][4]}}{[3]} &  \\ \\
& \frac{\sqrt{[2][4]}}{[3]}& \frac{1}{[3]} &   \\ \\ &&& 1 \end{array}\right),
\ \ \ \
\hat W_{[41]} =
\left(\begin{array}{cccc}   \\  -\frac{1}{[4]} & \frac{\sqrt{[3][5]}}{[4]}   \\ \\
\frac{\sqrt{[3][5]}}{[4]}& \frac{1}{[4]}  \\ \\ && 1 \\  &&& 1 \end{array}\right)
\ee
\be
\hat{\cal R}_{[32]} = \left(\begin{array}{ccccc} q \\ & q \\ && -\frac{1}{q}\\ &&& q\\&&&&-\frac{1}{q}\end{array}\right),
\ \ \ \
\hat U_{[32]} =
\left(\begin{array}{ccccc} 1 \\ &  -\frac{1}{[2]} & -\frac{\sqrt{[3]}}{[2]}& \\ \\
&   \frac{\sqrt{[3]}}{[2]}& -\frac{1}{[2]} \\ \\& & &-\frac{1}{[2]} & -\frac{\sqrt{[3]}}{[2]} \\ \\
& & & \frac{\sqrt{[3]}}{[2]}& -\frac{1}{[2]}\end{array}\right),
\ \ \ \ \nn \\ \nn \\
\hat V_{[32]} =
\left(\begin{array}{ccccc}   -\frac{1}{[3]} & \frac{\sqrt{[2][4]}}{[3]}   \\ \\
 \frac{\sqrt{[2][4]}}{[3]}& \frac{1}{[3]} \\  && 1 \\&&&1\\&&&&-1\end{array}\right),
\ \ \ \
\hat W_{[32]} =
\left(\begin{array}{ccccc}  1\\  &-\frac{1}{[2]} && \frac{\sqrt{[3]}}{[2]}   \\ \\ &&-\frac{1}{[2]} && \frac{\sqrt{[3]}}{[2]}   \\ \\
&\frac{\sqrt{[3]}}{[2]}&& \frac{1}{[2]}  \\ \\&&\frac{\sqrt{[3]}}{[2]}&& \frac{1}{[2]}   \end{array}\right)
\label{32}\ee
and
\be
\hat{\cal R}_{[311]} = \left(\begin{array}{cccccc} q \\ & q \\ && -\frac{1}{q}\\ &&& q\\&&&&-\frac{1}{q}\\&&&&&-\frac{1}{q}\end{array}\right),
\ \ \ \
\hat U_{[311]} =
\left(\begin{array}{cccccc} 1&&&&&\\
&-\frac{1}{[2]}&-\frac{\sqrt{[3]}}{[2]}\\ \\
&\frac{\sqrt{[3]}}{[2]}&-\frac{1}{[2]}\\
&&&-\frac{1}{[2]}&-\frac{\sqrt{[3]}}{[2]}\\ \\
&&&\frac{\sqrt{[3]}}{[2]}&-\frac{1}{[2]}\\ \\
&&&&&1\\\end{array}\right),
\ \ \ \ \nn \\ \nn
\ee
\be
\hat V_{[311]} =
\left(\begin{array}{cccccc}  -\frac{1}{[3]}&\frac{\sqrt{[2][4]}}{[3]}\\ \\
\frac{\sqrt{[2][4]}}{[3]}&\frac{1}{[3]}\\
&&1\\
&&&-1\\
&&&&-\frac{1}{[3]}&\frac{\sqrt{[2][4]}}{[3]}\\ \\
&&&&\frac{\sqrt{[2][4]}}{[3]}&\frac{1}{[3]}\\\end{array}\right),
\ \ \ \ \
\hat W_{[311]} =
\left(\begin{array}{cccccc}  -1\\
&-\frac{1}{[4]}&&\frac{\sqrt{[3][5]}}{[4]}\\ \\
&&-\frac{1}{[4]}&&\frac{\sqrt{[3][5]}}{[4]}\\ \\
&\frac{\sqrt{[3][5]}}{[4]}&&\frac{1}{[4]}\\ \\
&&\frac{\sqrt{[3][5]}}{[4]}&&\frac{1}{[4]}\\ \\
&&&&&1\\\end{array}\right)
\label{311}\ee

\subsection*{m=6}

\be
\boxed{
\begin{array}{c} \\
{\cal H}_{_\Box}^{(a_1b_1c_1d_1e_1|a_2b_2c_2d_2e_2|\ldots)}
= q^{\sum_i (a_i+b_i+c_i+d_i+e_i)} S_{[6]}+ (mirror) +
\\
\\
+ \sum_Q S_{Q} \cdot \Tr \prod_i \left(
\hat{\cal R}_{Q}^{a_i}\hat U_{Q} \hat{\cal R}_{Q}^{b_i}\hat V_{Q}\hat U_{Q}
\hat{\cal R}_{Q}^{c_i}\hat W_{Q}\hat V_{Q}\hat U_{Q} \hat{\cal R}_{Q}^{d_i}
\hat Y_{Q}\hat W_{Q}\hat V_{Q}\hat U_{Q}\hat{\cal R}_{Q}^{e_i}
\hat U_{Q}^\dagger\hat V_{Q}^\dagger\hat W_{Q}^\dagger\hat Y_{Q}^\dagger
\hat U_{Q}^\dagger\hat V_{Q}^\dagger\hat W_{Q}^\dagger
\hat U_{Q}^\dagger\hat V_{Q}^\dagger\hat U_{Q}^\dagger\right)
\end{array}
}
\label{6str}
\ee

\bigskip

\noindent
Here the sum goes over four representations $[51]$, $[42]$, $[411]$, $[33]$
plus the contributions of their mirrors $[21111]$, $[2211]$, $[3111]$, $[222]$
and of the symmetric diagram $Q=[321]$.
The contribution of $[111111]$ is explicitly mentioned in the first line
as a mirror of $[6]$: these both representations enter with
multiplicities one and no mixing matrices are involved.
All other relevant matrices are explicitly listed in Appendix 1.

\subsection*{m=7}

In this case, the formula looks exactly like (\ref{6str}), only one more piece with
$\hat{\cal R}_{Q}^{f_i}$ and $Z_Q$ is added:

\be
\boxed{
\begin{array}{c} \\
{\cal H}_{_\Box}^{(a_1b_1c_1d_1e_1f_1|a_2b_2c_2d_2e_2f_2|\ldots)} =
q^{\sum_i (a_i+b_i+c_i+d_i+e_i+f_i)} S_{[7]}+ (mirror) +
\\
\\
+ \sum_Q S_{Q} \cdot \Tr \prod_i \left(
\hat{\cal R}_{Q}^{a_i}\hat U_{Q} \hat{\cal R}_{Q}^{b_i}\hat V_{Q}\hat U_{Q}
\hat{\cal R}_{Q}^{c_i}\hat W_{Q}\hat V_{Q}\hat U_{Q} \hat{\cal R}_{Q}^{d_i}
\hat Y_{Q}\hat W_{Q}\hat V_{Q}\hat U_{Q}\hat{\cal R}_{Q}^{e_i}
\hat Z_{Q}\hat Y_{Q}\hat W_{Q}\hat V_{Q}\hat U_{Q}\hat{\cal R}_{Q}^{f_i}
\right.
\\
\\
\left.
\hat U_{Q}^\dagger\hat V_{Q}^\dagger\hat W_{Q}^\dagger\hat Y_{Q}^\dagger\hat Z_{Q}^\dagger
\hat U_{Q}^\dagger\hat V_{Q}^\dagger\hat W_{Q}^\dagger\hat Y_{Q}^\dagger
\hat U_{Q}^\dagger\hat V_{Q}^\dagger\hat W_{Q}^\dagger
\hat U_{Q}^\dagger\hat V_{Q}^\dagger\hat U_{Q}^\dagger\right)
\end{array}
}
\label{7str}
\ee
The mixing matrices are up to $20\times 20$, they are are explicitly listed in Appendix 2.

\section{Skein relations}

All matrices $\hat{\cal R}_Q$ have the same eigenvalues $\lambda=q$ and $\lambda=-\frac{1}{q}$, which both satisfy
\be
\lambda - \lambda^{-1} = q-q^{-1}
\ee
Therefore, each matrix satisfies
\be
\hat{\cal R}_Q - \hat{\cal R}_Q^{-1} = \left(q - \frac{1}{q}\right)\cdot I_Q
\ee
This implies simple difference equations, which for $R=[1]$ looks the same
for all coefficients (\ref{coefHOMFLY}) and, hence,
the extended HOMFLY polynomial (\ref{extHOMFLY}),
considered as a function of any of the braid parameters $a_{ij}$
in (\ref{brapa}) satisfies
\be
{\cal H}_{_\Box}^{(\ldots\ a_{ij}+1\ \ldots)}\{p\,\}
- {\cal H}_{_\Box}^{(\ldots\ a_{ij}-1\ \ldots)}\{p\,\}
= \left(q - \frac{1}{q}\right){\cal H}_{_\Box}^{(\ldots\ a_{ij}\ \ldots)}\{p\,\}
\ee
Note that this skein relation holds for the {\it extended} HOMFLY polynomials,
with arbitrary values of time variables $\{p_k\}$ beyond the topological locus
(\ref{tolo}), for arbitrary parameters $a_{ij}$,
but only for the fundamental representation $R=\Box=[1]$.
If complemented by the symmetry condition
\be
{\cal H}_{_\Box}^{-{\cal B}}\bigl(q\big|\{p\,\}\bigr)=
{\cal H}_{_\Box}^{{\cal B}}\bigl(q^{-1}\big|\{p\,\}\bigr)
\ee
this skein relation can be used to recursively find all the HOMFLY
polynomials for all braids.
The ordinary skein relation \cite{skein} for the HOMFLY polynomials $H_{_\Box}^{\cal K}(A|q)$
on the topological locus follows immediately:
\be
A\cdot{H}_{_\Box}^{(\ldots\ a_{ij}+1\ \ldots)}(q|A)
- \frac{1}{A}\cdot {H}_{_\Box}^{(\ldots\ a_{ij}-1\ \ldots)}(q|A)
= \left(q - \frac{1}{q}\right){H}_{_\Box}^{(\ldots\ a_{ij}\ \ldots)}(q|A)
\ee
where the extra powers of $A$ at the l.h.s. arise from the correcting factor in (\ref{framing})
(note that $|\Box| = 1$ and $\varkappa_{_\Box} = 0$).

\section{Applications of the 5-, 6- and 7-strand formulas}

Eq.(\ref{311}) is the only new one in the 5-strand case as compared to \cite{ammII},
but it allows one to study arbitrary 5-strand knots.
We consider two applications:
to the torus knots and to all $5$-strand knots with 9 crossings listed in \cite{katlas} (see the Tables below).

Similarly, with eqs.(\ref{6str}) and (\ref{7str}) we list all $6$-strand knots with 10 crossings listed in \cite{katlas}
(also see the Tables below). For an illustrative purpose we also list in the Tables an example of answer for
a $7$-strand knot from \cite{katlas}.

\subsection{Torus knots $[5,n]$}

For the torus knots and links $a_{ij}=1$ for $i\leq n$ and all other $a_{ij}=0$. These knots/links are denoted
$T[m,n]$, where $m$ is the number of strand (it is a knot if $m$ and $n$ are mutually prime and is an $l$-component
link if $l$ is the greatest common divisor of $m$ and $n$). In particular, for $m=5$ this means that (\ref{5str}) reduces to
\be
{\cal H}_{_\Box}^{[5,n]} = \sum_{Q\vdash 5}  S_Q\cdot
\Tr_{_{N_{1Q}\times N_{1Q}}} \Big(
\underbrace{\hat{\cal R}_Q\ \hat U_Q\hat{\cal R}_Q \hat V_Q \hat U_Q\hat{\cal R}_Q\hat W_Q \hat V_Q\hat U_Q\hat{\cal R}_Q
\hat U_Q^\dagger \hat V_Q^\dagger \hat U_Q^\dagger \hat W_Q^\dagger\hat V_Q^\dagger\hat U_Q^\dagger}_{{\mathfrak{R}}_Q}\Big)^n
\ee
From the explicit form of the constituent matrices it is easy to find the eigenvalues of
the composite matrices ${{\mathfrak{R}}_Q}$:
\be
\det_{_{N_{1Q}\times N_{1Q}}}\Big({{\mathfrak{R}}_Q} - \lambda\cdot I\Big) =
\left\{\begin{array}{ccc}
\cfrac{\lambda^5-q^{10}}{\lambda-q^2}& {\rm for} & Q=[41] \\ \\
\lambda^5-q^4& {\rm for} & Q=[32] \\ \\
(\lambda-1)(\lambda^5-1) & {\rm for} & Q=[311]\\ \\
\lambda^5-q^{-4} & {\rm for} & Q=[221] \\ \\
\cfrac{\lambda^5-q^{-10}}{\lambda-q^{-2}}& {\rm for}  & Q=[2111] \\
\end{array}\right.
\ee
This implies that for $n \not\vdots\ 5$, i.e. for the torus knots,
\be\label{1}
{\cal H}_{_\Box}^{[5,n]} = q^{4n} S_{[5]} -q^{2n} S_{[41]} + S_{[311]} - q^{-2n} S_{[2111]}
+ q^{-4n} S_{[11111]}
\ee
and for $ n\ \vdots\ 5$, i.e. for the torus 5-component links
\be\label{2}
{\cal H}_{_\Box}^{[5,n]} = q^{4n}S_{[5]} + 4q^{2n}S_{[41]} + 5q^{\frac{4n}{5}}S_{[32]} + 6S_{[311]}
+ 5q^{-\frac{4n}{5}}S_{[221]}
+ 4q^{-2n}S_{[2111]} + q^{-4n}S_{[11111]}
\ee
in full accordance with the Adams rule of \cite{chi}\footnote{The Adams rule consists of
constructing a set of coefficients
$C^Q_{\{R_i\}}$ via the decomposition of the product
\be
\prod_{i=1}^l S_{R_i}(p_k^{(m)})=\sum_{Q\vdash m\sum_iR_i} C^Q_{\{R_i\}}S_Q(p_k)
\ee
where $p_k^{(m)}\equiv p_{mk}$, $m$ and $l$ are integers and the sign $\vdash$ implies that the sum runs over
all representations of the size $m\sum_i|R_i|$. With these coefficients, the
colored HOMFLY polynomial for the
torus link $T[m,n]$ with $l$ components (i.e. $l$ is the greatest common divisor of $m$ and $n$) is
\be
{\cal H}_{_{\{R_i\}}}^{[5,n]}=\sum_{Q\vdash m\sum_iR_i} q^{{2n\over m}\varkappa_Q}C^Q_{\{R_i\}}S_Q(p_k)
\ee
}.

In order to obtain ${H}_{_\Box}^{[5,n]}$, one now has to multiply (\ref{1}) and (\ref{2}) by $A^{-4n}$ in
accordance with (\ref{framing}) and use values (\ref{Schurtolo}) of the Schur functions at the topological locus
(\ref{tolo}). These are
\be
S_{[5]}(p^*)={\{A\}\{Aq\}\{Aq^2\}\{Aq^3\}\{Aq^4\}\over \{q\}\{q^2\}\{q^3\}\{q^4\}\{q^5\}}\ \ \ \ \ \ \ \ \  \
S_{[41]}(p^*)={\{A/q\}\{A\}\{Aq\}\{Aq^2\}\{Aq^3\}\over \{q\}^2\{q^2\}\{q^3\}\{q^5\}}\\
S_{[311]}(p^*)={\{A/q^2\}\{A/q\}\{A\}\{Aq\}\{Aq^2\}\over \{q\}^2\{q^2\}^2\{q^5\}}\ \ \ \ \ \ \ \ \  \
S_{[32]}(p^*)={\{A/q\}\{A\}^2\{Aq\}\{Aq^2\}\over \{q\}^2\{q^2\}\{q^3\}\{q^4\}}\\
S_{[2111]}(p^*)={\{Aq\}\{A\}\{A/q\}\{A/q^2\}\{A/q^3\}\over \{q\}^2\{q^2\}\{q^3\}\{q^5\}}\ \ \ \ \ \ \ \ \  \
S_{[221]}(p^*)={\{Aq\}\{A\}^2\{A/q\}\{A/q^2\}\over \{q\}^2\{q^2\}\{q^3\}\{q^4\}}\\
S_{[11111]}(p^*)={\{A\}\{A/q\}\{A/q^2\}\{A/q^3\}\{A/q^4\}\over \{q\}\{q^2\}\{q^3\}\{q^4\}\{q^5\}}
\ee

\subsection{Knots and links from the katlas tables}

The 5-strand and 6-strand formulas ({\ref{5str}}) and (\ref{6str})
enable to calculate the HOMFLY polynomials
for arbitrary 5- and 6-strand knots straightforwardly.
Coefficients of the character expansion for the HOMFLY polynomials
for all the 5-strand knots with 9 crossings and for all 6-strand knots with 10 crossings are given in the Tables below.
We also give there coefficients of this expansion for 7-strand knot $12a_{0125}$.

\subsection{Cabling of $2$ and $3$-strand links}

Like it was done in \cite{iammIII}, one can now reproduce various
formulas for the colored HOMFLY polynomials by the {\it cabling} procedure
applied to the fundamental representations. Namely, the $5$-strand
links can describe the 2-strand links, where one component is in
$[2]$ or $[11]$, and the other one is in $[3]$, $[21]$ or $[111]$.
Similarly, the $6$-strand knots can be used for cabling of the
$2$-strand knots in representation $[3],\ [21], \ [111]$ and of the
$3$-strand knots in representations $[2]$ or $[11]$. The last option
is most interesting, because it tests the $3$-strand formulas of
\cite{iammIII}, the simplest case of non-torus knots, where the
Rosso-Jones formula \cite{chi} is unapplicable.

In this case, the projectors from the cable of the knot to irrep are
(see also \cite{iammIII}; the projectors are definitely not uniquely defined,
these were just the simplest ones
made out of zero and one insertions of $\hat{\cal R}$)
\be
{\cal H}_{[2]}^{(a_1b_1|a_2b_2|\ldots)}\{p\,\} =
\frac{1}{(1+q^2)^3}\sum_{i,j,k=0,1}q^{i+j+k} {\cal H}_{_\Box}^{\big(i0j0k\big|
(01100|11000)^{a_1}\big|(00011|00110)^{b_1}
\big|(01100|11000)^{a_2}
\big|\ldots\big)}\{p\,\}
\nn \\
\\
{\cal H}_{[11]}^{(a_1b_1|a_2b_2|\ldots)}\{p\,\} =
\frac{q^6}{(1+q^2)^3}\sum_{i,j,k=0,1}  \left(-\frac{1}{q}\right)^{i+j+k}
{\cal H}_{_\Box}^{\big(i0j0k\big|
(01100|11000)^{a_1}\big|(00011|00110)^{b_1}
\big|(01100|11000)^{a_2}
\big|\ldots\big)}\{p\,\}
\nn
\ee
The braiding at the r.h.s. can be directly read from the cabling picture (these are elementary braiding patterns
in terms of the integers $a_{ij}$ for the 6-strand braid, i.e., for instance, $(00011|00110)$ means
that first the 4th and 5th strands cross, then the 5th and 6th, the 3d and 4th and, finally, the 4th and 5th ones):

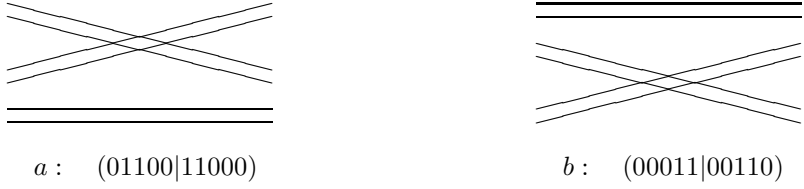
\begin{picture}(20,20)(-55,25)
\put(0,0){\line(4,1){100}}
\put(0,5){\line(4,1){100}}
\put(0,-10){\line(1,0){100}}
\put(0,-15){\line(1,0){100}}
\put(0,30){\line(4,-1){100}}
\put(0,25){\line(4,-1){100}}
\put(200,-10){\line(4,1){100}}
\put(200,-15){\line(4,1){100}}
\put(200,30){\line(1,0){100}}
\put(200,25){\line(1,0){100}}
\put(200,10){\line(4,-1){100}}
\put(200,15){\line(4,-1){100}}
\put(10,-35){\mbox{$a: \ \ \ (01100|11000)$}}
\put(210,-35){\mbox{$b: \ \ \ (00011|00110)$}}
\end{picture}
\vspace{2.3cm}\\
Making use of the 6-strand formula (\ref{6str}) for the r.h.s. and
equally explicit expressions for the colored 3-strand braids from \cite{iammIII}
for the l.h.s.,
one can easily check that these equalities are indeed true
for various choices of the braiding numbers $a_1,b_1,a_2,b_2,a_3,\ldots$.

\bigskip

As to the $5$-strand example, the HOMFLY polynomials for the bi-colored $2$-strand links
${\cal H}_{[2][3]}^{(a_1|a_2|a_3|\ldots)}\{p\,\}$ with {\it even} number of
parameters $a_1,a_2,\ldots = \pm 1$ can be evaluated with the help of
the Rosso-Jones formula \cite{chi}, it is valid also for the {\it extended}
HOMFLY polynomials.
On the other hand, it can be extracted by cabling from the $5$-strand formula (\ref{5str}).
Thus,

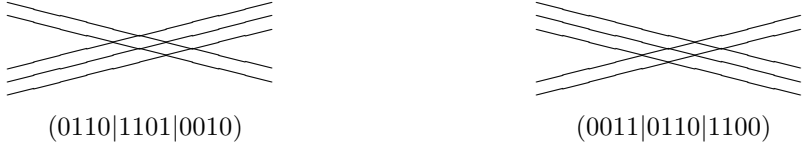
\begin{picture}(20,20)(-55,25)
\put(0,0){\line(4,1){100}}
\put(0,5){\line(4,1){100}}
\put(0,10){\line(4,1){100}}
\put(0,30){\line(4,-1){100}}
\put(0,35){\line(4,-1){100}}
\put(200,0){\line(4,1){100}}
\put(200,5){\line(4,1){100}}
\put(200,25){\line(4,-1){100}}
\put(200,30){\line(4,-1){100}}
\put(200,35){\line(4,-1){100}}
\put(15,-15){\mbox{$(0110|1101|0010)$}}
\put(215,-15){\mbox{$(0011|0110|1100)$}}
\end{picture}
\vspace{1.7cm}
\be
{\cal H}_{[2][3]}^{(a_1|a_2|a_3|\ldots)}\{p\,\} =
\sum_{i,j,k}\ \pi_{[2]}^i\pi_{[3]}^{jk}\ {\cal H}_{_\Box}^{\big(i0jk\big|
(0110|1101|0010)^{a_1}\big|(0011|0110|1100)^{a_2}\big|
(0110|1101|0010)^{a_3}
\big|\ldots\big)}\{p\,\}
\ee
The projectors $\pi$ are derived as follows.

From \cite{iammIII}, the projector on $[2]$ is
$\pi_{[2]}^i = \frac{q^i}{1+q^2}$ with $i=0,1$.
That on $[11]$ is just its mirror $\pi_{[11]}^i = \frac{(-)^iq^{2-i}}{1+q^2}$.
This follows from the relation
\begin{picture}(20,20)(50,25)
\put(0,0){\line(1,0){30}}
\put(0,10){\line(1,0){30}}
\put(80,0){\line(4,1){40}}
\put(80,10){\line(4,-1){40}}
\put(-10,3){\mbox{$\alpha$}}
\put(48,3){\mbox{$+\ \ \ \beta$}}
\put(-50,3){\mbox{$\pi_{[2]}\ = $}}
\end{picture}
\be
\ \ \ \ \ \ \ \ \ \ \ \ \ \ \ \ \ \ \ \ \
\ \ \ \ \ \ \ \ \ \ \ \ \ \ \ \ \ \ \ \ \
\ \ \ \ \ \ \ \ \ \ \ \ \ \ \ \ \ \ \ \ \
S_{[2]}\{p\,\} = \alpha(S_{[2]}+S_{[11]}) + \beta(qS_{[2]}-q^{-1}S_{[11]})
\ee
which implies that
$\alpha = \frac{1}{q[2]} = \frac{1}{1+q^2}$ and $\beta= \frac{1}{[2]} = \frac{q}{1+q^2}$.\\

Similarly, for $\pi_{[3]}$ one can make a choice
\begin{picture}(20,20)(20,25)
\put(0,0){\line(1,0){30}}
\put(0,10){\line(1,0){30}}
\put(0,-10){\line(1,0){30}}
\put(80,0){\line(4,1){40}}
\put(80,10){\line(4,-1){40}}
\put(80,-10){\line(1,0){40}}
\put(-10,-2){\mbox{$\alpha$}}
\put(48,-2){\mbox{$+\ \ \ \beta$}}
\put(-50,-2){\mbox{$\pi_{[3]}\ = $}}
\put(130,-2){\mbox{$+\ \ \ \gamma$}}
\put(160,0){\line(4,1){60}}
\put(160,10){\line(4,-1){60}}
\put(160,-10){\line(4,1){60}}
\end{picture}
\vspace{1.3cm}
\be
S_{[3]}\{p\,\} = \alpha \Big(S_{[3]} + 2S_{[21]} + S_{[111]}\Big)
+ \beta \Big(qS_{[3]} + (q-q^{-1})S_{[21]} - q^{-1} S_{[111]}\Big)
+ \gamma \Big(q^2S_{[3]} - S_{[21]} + q^{-2}S_{[111]}\Big)
\ee
so that $\alpha = \frac{1}{q^2[2][3]}$, $\beta = \frac{q^2+2}{q^2[2][3]}$, $\gamma = \frac{1}{[3]}$
and
\be
\pi_{[3]}^{jk} = \frac{1}{q^2[2][3]}\Big(\delta^{j0}\delta^{k0} + (2+q^2)\delta^{j1}\delta_{k0}
+q^2[2]\delta^{i1}\delta^{j1}\Big)
\ee
Likewise for the projection on the mirror representation $[111]$, one gets
$\alpha = \frac{q^2}{[2][3]}$, $\beta = -\frac{q^2+2}{q^2[2][3]}$, $\gamma = \frac{1}{[3]}$
and for the projection on $[21]$:
$\alpha = \frac{1}{[3]}$, $\beta = \frac{q-q^{-1}}{[3]}$, $\gamma = -\frac{1}{[3]}$.
Once again, there is a big freedom in choosing the projectors, and this is just
one possibility, not distinguished in any way.

In the same way, one can use the $6$- and $7$-strand formulas from this paper
to represent the $2$-strand links, bi-colored as $[2][4]$, $[2][5]$, $[3][3]$ and $[3][4]$.

\section{Conclusion}

To conclude, we represented the extended HOMFLY polynomials
(Wilson loop averages in $3d$ Chern-Simons theory)
in the fundamental representation
for arbitrary $5$-, $6$- and $7$-strand braids as linear combinations
of respectively seven, eleven and fifteen Schur functions
with $q$-dependent coefficients,
which are traces of at most $6\times 6$, $16\times 16$ and $20\times 20$
explicitly listed matrices.
Parameters (\ref{brapa}) of the braid enter as powers
of the diagonal constituents of these matrices.
These formulas immediately reproduce all the $5$-, $6$- and $7$-strand formulas
in \cite{katlas}.
Application to torus knots and links $[5,n]$ in s.6.1 illustrates
the possibility to describe arbitrary {\it series} of
knots\!/links by evaluating the eigenvalues of the corresponding
composite ${\cal R}$-matrices.

\bigskip

Actually, the result of this paper can be far more ambitious:
if the simple rule eq.(\ref{Rc}) (which does not depend on $q$ and, hence,
can be read off not only from the quantum groups, but from the Lie groups as well) remains true
beyond $SU_q(3)$, then we provided a {\it complete}
description of the fundamental HOMFLY polynomials
for {\it all} braids with any number $m$ of strands.
Moreover, this description provides an explicit function
of the braiding numbers, and not only for ordinary,
but for the {\it extended} HOMFLY polynomials as well.

Given the importance of this result, we summarize
briefly the steps of the construction.
\begin{itemize}
\item The Turaev-Reshetikhin description of the
HOMFLY polynomials is projected on the space of intertwining
operators which directly provides their character expansion
naturally continued to the infinite-dimensional
space of the time variables $\{p_k\}$.
\item The expansion coefficients are traces of products
of the diagonal $\hat{\cal R}$-matrices with generically
known entries dictated by the eigenvalues of the
cut-and-join operator $W_{[2]}$
(certain symmetric group characters)
and the orthogonal mixing matrices.
\item The mixing matrices are products of the Racah matrices.
\item For the fundamental representation the Racah matrices
are made from the $2\times 2$ blocks; this decomposition
is immediately read off from {\it the representation tree}
and can be easily programmed.
\item The cosines of angles of the $2\times 2$
rotation matrices are inverse quantum integers.
These integers are the only non-trivial parameters to
calculate.
\item An explicit calculation of the Racah coefficients for
$SU_q(2)$ and $SU_q(3)$ quantum groups
provides a formula (\ref{Rc}) for these integers,
which is sufficient to describe all the $6$-strand braids.
\item For the hook diagrams the answer for the integers
is clearly universal, this allows us to conjecture
the answer for representation $[4111]$, the only one
appearing in the $7$-strand formulas, which needs
$SU_q(4)$ for being honestly calculated.
\item In fact, eq.(\ref{Rc}) looks so nice that it
can easily remain true for all bigger quantum groups
and thus provides an answer for braids with arbitrary
number of strands. If this is the case, one gets
{\bf a complete description of {\it all} extended
HOMFLY polynomials for all braids}.
\item The only restriction is to the fundamental
representation. However, since the formulas are true for {\it any}
braids, the cabling procedure is actually straightforward
as demonstrated in sec.6.3. Still, the formulas
obtained by cabling seem more complicated than the direct
counterparts of our results in the fundamental case,
and development of a similar formalism for colored knots
is highly desirable (see \cite{iammIII,IMMMfe} for the
first results in this direction).
\end{itemize}

\section*{Acknowledgements}

Our work is partly supported by Ministry of Education and Science of
the Russian Federation under contracts 14.740.11.0347 (A.A. and And.Mor.)
and 14.740.11.0347 (A.Mir. and A.Mor.), by NSh-3349.2012.2,
by RFBR grants 10-01-00836-a (A.A. and And.Mor.) and 10-01-00536 (A.Mir. and A.Mor.) and
by joint grants 11-02-90453-Ukr, 12-02-91000-ANF,
11-01-92612-Royal Society, 12-02-92108-Yaf-a.

\newpage
\thispagestyle{empty}

\voffset=-1.6in

{\tiny{\rotate{
\begin{tabular}{|c|cccccc|}
\multicolumn{7}{c}{{\small\bf The Table of HOMFLY polynomials for all 5-strand knots with 9 crossings}}\\ \multicolumn{7}{c}{}\\ \hline
&&&&&&\\
{\rm knot} &$(a_1,b_1,c_1,d_1|a_2,b_2,c_2,d_2|a_3,b_3,c_3,d_3,\ldots)$ &
$S_5^*$& \multicolumn{2}{c}{$S_{41}^*$} & \multicolumn{2}{c|}{$S_{32}^*$}\\
&&&&&&\\ \hline
&&&&&&\\
$9_2$&$(-3, -1, 0, 0|1, -1, -1, 0|0, 1, -1, -1|0, 0, 1, -1)$&
${q}^{-6}$&\multicolumn{2}{c}{$-{q}^{2}+4-7\,{q}^{-2}+6\,{q}^{-4}-3\,{q}^{-6}$}&
\multicolumn{2}{c|}{$-{q}^{8}+{q}^{6}-2\,{q}^{2}+6-9\,{q}^{-2}+8\,{q}^{-4}-5\,{q}^{-6}+2\,{
q}^{-8}
$}
\\
&&&&&&\\
\hline
\end{tabular}
}
}}

{\tiny\rotate{
\begin{tabular}{|c|cccc|}
\multicolumn{5}{c}{{\normalsize \bf The Table of HOMFLY polynomials for all $6$-strand knots with $10$ crossings}}
\\\multicolumn{5}{c}{}\\
\multicolumn{5}{c}{\normalsize The contribution of the remaining diagrams is obtained with help of mirror symmetry}
\\\multicolumn{5}{c}{}\\
\multicolumn{5}{c}{\normalsize $q\ \mapsto\ -\frac{1}{q},\ \ \ \ S^*_6\ \mapsto\ S^*_{111111},\ \ \ S^*_{51}\ \mapsto\ S^*_{21111},\ \ \ S^*_{42}\ \mapsto\ S^*_{2211},\ \ \ S^*_{411}\ \mapsto\ S^*_{3111},\ \ \ S^*_{33}\ \mapsto\ S^*_{222}$}\\\multicolumn{5}{c}{}\\
\hline&&&&\\
{\rm knot}&$(a_1,b_1,c_1,d_1,e_1|a_2,b_2,c_2,d_2,e_2|a_3,b_3,c_3,d_3,e_3\ldots)$&$S^*_6$&$S^*_{51}$&$S^*_{42}$\\
\hline
\end{tabular}}}
\newpage
\begin{center}
{\bf The HOMFLY polynomial for an example of $7$-strand knot}
\\The contribution of the remain diagrams is obtained with help of mirror symmetry\\$
q\ \mapsto\ \frac{1}{q},\ \ \ \ S^*_{7}\ \mapsto\ S^*_{1111111},\ \ \ S^*_{61}\ \mapsto\ S^*_{211111},\ \ \ S^*_{52}\ \mapsto\ S^*_{22111},\ \ \ S^*_{511}\ \mapsto\ S^*_{31111},$\\$ S^*_{43}\ \mapsto\ S^*_{2221},\ \ \ S^*_{421}\ \mapsto\ S^*_{3211},\ \ \ S^*_{331}\ \mapsto\ S^*_{322}$
\end{center}
$$
\begin{array}{|c|c|}
\hline
\multicolumn{2}{|c|}{{\rm knot}\ 12a_{0125}}\\\hline\multicolumn{2}{|c|}{}\\
\multicolumn{2}{|c|}{(a_1,b_1,c_1,d_1,e_1,f_1|a_2,b_22,c_2,d_2,e_2,f_2|a_3,b_3,c_3,d_3,e_3,f_3)}\\\multicolumn{2}{|c|}{}\\
\multicolumn{2}{|c|}{(1, -1, -1, 0, 0, 0|1, 0, -1, -1, 1, -1|0, -1, 1, 2, 1, -1)}\\\multicolumn{2}{|c|}{}\\\hline&\\
S^*_7&1\\&\\
S^*_{61}&{q}^{6}-5\,{q}^{2}+7-5\,{q}^{-2}+{q}^{-6}\\&\\
S^*_{52}&{q}^{10}-3\,{q}^{8}+16\,{q}^{4}-41\,{q}^{2}+54-41\,{q}^{-2}+16\,{q}^{-
4}-3\,{q}^{-8}+{q}^{-10}
\\&\\
S^*_{511}&{q}^{10}-2\,{q}^{8}-6\,{q}^{6}+23\,{q}^{4}-35\,{q}^{2}+39-35\,{q}^{-2}
+23\,{q}^{-4}-6\,{q}^{-6}-2\,{q}^{-8}+{q}^{-10}
\\&\\
S^*_{43}&-{q}^{10}+7\,{q}^{8}-25\,{q}^{6}+56\,{q}^{4}-89\,{q}^{2}+104-89\,{q}^{
-2}+56\,{q}^{-4}-25\,{q}^{-6}+7\,{q}^{-8}-{q}^{-10}
\\&\\
S^*_{421}&{q}^{12}-9\,{q}^{10}+35\,{q}^{8}-82\,{q}^{6}+135\,{q}^{4}-174\,{q}^{2}
+188-174\,{q}^{-2}+135\,{q}^{-4}-82\,{q}^{-6}+35\,{q}^{-8}-9\,{q}^{-10
}+{q}^{-12}
\\&\\
S^*_{4111}&{q}^{12}-9\,{q}^{10}+29\,{q}^{8}-48\,{q}^{6}+56\,{q}^{4}-62\,{q}^{2}+
65-62\,{q}^{-2}+56\,{q}^{-4}-48\,{q}^{-6}+29\,{q}^{-8}-9\,{q}^{-10}+{q
}^{-12}
\\&\\
S^*_{331}&{q}^{12}-8\,{q}^{10}+29\,{q}^{8}-66\,{q}^{6}+112\,{q}^{4}-151\,{q}^{2}
+166-151\,{q}^{-2}+112\,{q}^{-4}-66\,{q}^{-6}+29\,{q}^{-8}-8\,{q}^{-10
}+{q}^{-12}
\\&\\
\hline\multicolumn{2}{c}{}
\end{array}
$$

\newpage
\voffset=-1.1in

\section*{Appendix 1. Mixing and ${\cal R}$-matrices for the 6-strand formula
(\ref{6str})}

{\bf $5\times 5$ mixing matrices for $[5,1]$}
\be
\hat\hotR_{[5,1]}=\small{\left(\begin{array}{ccccc}\\q\\&q\\&&q\\&&&q\\ &&&&-\cfrac{1}{q}\\  \\ \end{array}\right)},\ \ \ \ \
\hat U_{[5,1]}=\footnotesize{\left(\begin{array}{ccccc}\\1\\&1\\&&1\\ &&&-\cfrac{1}{ [2]}&\cfrac{\sqrt{[3]}}{\ \ [2]}\\ \\&&&-\cfrac{\sqrt{[3]}}{[2]}&-\cfrac{1}{[2]}\\  \\ \end{array}\right)},\ \ \ \
\hat V_{[5,1]}=\footnotesize{\left(\begin{array}{ccccc}\\1\\&1\\&&-\cfrac{1}{[3]}&\cfrac{\sqrt{[2][4]}}{[3]}\\ \\&&\cfrac{\sqrt{[2][4]}}{[3]}&\cfrac{1}{[3]}\\  \\&&&&1\\  \\\end{array}\right)},  \nn \\ \nn\\ \nn \\\nn\\
\hat W_{[5,1]}=\footnotesize{\left(\begin{array}{ccccc}\\1\\&-\cfrac{1}{[4]}&\cfrac{\sqrt{[3][5]}}{[4]}\\ \\&\cfrac{\sqrt{[3][5]}}{[4]}&\cfrac{1}{[4]}\\ \\&&&1\\&&&&1\\ \\ \end{array}\right)},\ \ \ \
\hat Y_{[5,1]}=\footnotesize{\left(\begin{array}{ccccc}\\-\cfrac{1}{[5]}&\cfrac{\sqrt{[4][6]}}{[5]}\\ \\ \cfrac{\sqrt{[4][6]}}{[5]}&\cfrac{1}{[5]}\\ \\&&1\\&&&1\\&&&&1\\  \\ \end{array}\right)}
\nn\ee

\bigskip\bigskip
\bigskip\bigskip
\bigskip\bigskip
\bigskip\bigskip

{\bf $5\times 5$ mixing matrices for $[3,3]$}
\be
\hspace{-1.8cm}\hat\hotR_{[3,3]}=\small{\left(\begin{array}{ccccc}\\q\\&q\\&&-\cfrac{1}{q}\\&&&q\\&&&&-\cfrac{1}{q}\\ \\ \end{array}\right)},\ \ \ \
\hat U_{[3,3]}=\footnotesize{\left(\begin{array}{ccccc}\\1\\&-\cfrac{1}{[2]}&\cfrac{\sqrt{[3]}}{[2]}\\ \\&-\cfrac{\sqrt{[3]}}{[2]}&-\cfrac{1}{[2]}\\ \\&&&-\cfrac{1}{[2]}&\cfrac{\sqrt{[3]}}{[2]}\\ \\&&&-\cfrac{\sqrt{[3]}}{[2]}&-\cfrac{1}{[2]}\\  \\\end{array} \right)},\ \ \ \
\hat V_{[3,3]}=\footnotesize{\left(\begin{array}{ccccc}-\cfrac{1}{[3]}&\cfrac{\sqrt{[2][4]}}{[3]}\\ \\ \cfrac{\sqrt{[2][4]}}{[3]}&\cfrac{1}{[3]}\\\\&&1\\&&&1\\ &&&&-1\\  \\ \end{array}\right)}, \nn\\ \nn \\ \nn \\ \nn \\
\hat W_{[3,3]}=\footnotesize{\left(\begin{array}{ccccc}\\1\\&-\cfrac{1}{[2]}&&\cfrac{\sqrt{[3]}}{[2]} \\&&-\cfrac{1}{[2]}&&\cfrac{\sqrt{[3]}}{[2]} \\&\cfrac{\sqrt{[3]}}{[2]}&&\cfrac{1}{[2]}\\ \\ &&\cfrac{\sqrt{[3]}}{[2]}&&\cfrac{1}{[2]} \\ \\\end{array}\right)},\ \ \ \
\hat Y_{[3,3]}=\small{\left(\begin{array}{ccccc}\\ 1 \\ &1 \\&&1\\ &&&-1\\&&&&-1\\ \\ \end{array}\right)}
\nn\ee

\newpage

{\bf $9\times 9$ mixing matrices for $[4,2]$}
\be
\hspace{-1.5cm}\hat\hotR_{[4,2]}=\small{\left(\begin{array}{cccccccccc}\\q\\&q\\&&q\\&&&-\cfrac{1}{q}\\&&&&q\\&&&&&q\\&&&&&&-\cfrac{1}{q}
\\&&&&&&&q\\&&&&&&&&-\cfrac{1}{q}\\  \\\end{array}\right)},\
\hat U_{[4,2]}={\footnotesize\left(\begin{array}{ccccccccc}\\1\\&1\\ \\ &&-\cfrac{1}{[2]}&\cfrac{\sqrt{[3]}}{[2]}\\ \\
&&-\cfrac{\sqrt{[3]}}{[2]}&-\cfrac{1}{[2]}\\ \\ &&&&1\\ &&&&&-\cfrac{1}{[2]}&\cfrac{\sqrt{[3]}}{[2]}\\
\\&&&&&-\cfrac{\sqrt{[3]}}{[2]}&-\cfrac{1}{[2]}\\  \\ &&&&&&&-\cfrac{1}{[2]}&\cfrac{\sqrt{[3]}}{[2]}\\
\\&&&&&&&-\cfrac{\sqrt{[3]}}{[2]}&-\cfrac{1}{[2]}\\\end{array}\right)},\nn\\ \nn\\ \nn\\ \nn\\
 \!\!\!\!\hat V_{[4,2]}={\footnotesize\left(\begin{array}{cccccccccc}\\1\\&-\cfrac{1}{[3]}&\cfrac{\sqrt{[2][4]}}{[3]}\\
 \\&\cfrac{\sqrt{[2][4]}}{[3]}&\cfrac{1}{[3]}\\ \\ &&&1\\&&&&-\cfrac{1}{[3]}&\cfrac{\sqrt{[2][4]}}{[3]}\\
 \\&&&&\cfrac{\sqrt{[2][4]}}{[3]}&\cfrac{1}{[3]} \\  \\&&&&&&1\\&&&&&&&1\\&&&&&&&&-1 \\  \\ \end{array}\right)},\nn
\\
\hat W_{[4,2]}={\footnotesize\left(\begin{array}{cccccccccc}\\-\cfrac{1}{[4]}&\cfrac{\sqrt{[3][5]}}{[4]}\\
\\ \cfrac{\sqrt{[3][5]}}{[4]}&\cfrac{1}{[4]}\\  \\&&1\\&&&1\\&&&&1\\ \\ &&&&&-\cfrac{1}{[2]}&&\cfrac{\sqrt{[3]}}{[2]}
\\&&&&&&-\cfrac{1}{[2]}&&\cfrac{\sqrt{[3]}}{[2]}\\ &&&&&\cfrac{\sqrt{[3]}}{[2]}&&\cfrac{1}{[2]} \\
&&&&&& \cfrac{\sqrt{[3]}}{[2]} &&\cfrac{1}{[2]}\\ \\\end{array}\right)}, \nn\\\nn\\ \nn\\ \nn\\ \nn
\ee
\be
\hat Y_{[4,2]}={\footnotesize\left(\begin{array}{cccccccccc}\\1\\ \\&-\cfrac{1}{[3]}&&&\cfrac{\sqrt{[2][4]}}{[3]}\\
&&-\cfrac{1}{[3]}&&&\cfrac{\sqrt{[2][4]}}{[3]}\\&&&-\cfrac{1}{[3]}&&&\cfrac{\sqrt{[2][4]}}{[3]}\\
&\cfrac{\sqrt{[2][4]}}{[3]}&&&\cfrac{1}{[3]}\\ &&\cfrac{\sqrt{[2][4]}}{[3]}&&&\cfrac{1}{[3]}\\
&&&\cfrac{\sqrt{[2][4]}}{[3]}&&&\cfrac{1}{[3]}\\ \\&&&&&&&1\\&&&&&&&&1\\  \\ \end{array}\right)}\nn,
\ee

\bigskip

{\bf $10\times 10$ mixing matrices for $[4,1,1]$}
\be
\hspace{-1.5cm}\hat\hotR_{[4,1,1]}=\footnotesize{\left(\begin{array}{ccccccccccc}\\q\\&q\\&&q\\&&&-\cfrac{1}{q}\\&&&&q\\&&&&&q\\&&&&&&-\cfrac{1}{q}\\&&&&&&&q\\ \\
&&&&&&&&-\cfrac{1}{q}\\&&&&&&&&&-\cfrac{1}{q} \\ \\\end{array}\right)},\ \ \ \
\hat U_{[4,1,1]}=\scriptsize{\left(\begin{array}{ccccccccccc}\\1\\&1\\&&-\cfrac{1}{[2]}&\cfrac{\sqrt{[3]}}{[2]}\\ \\ &&-\cfrac{\sqrt{[3]}}{[2]}&-\cfrac{1}{[2]}\\ \\ \\&&&&1\\&&&&&-\cfrac{1}{[2]}&\cfrac{\sqrt{[3]}}{[2]}\\ \\&&&&&-\cfrac{\sqrt{[3]}}{[2]}&-\cfrac{1}{[2]}\\&&&&&&&-\cfrac{1}{[2]}&\cfrac{\sqrt{[3]}}{[2]}\\ \\ &&&&&&&-\cfrac{\sqrt{[3]}}{[2]}&-\cfrac{1}{[2]}\\ \\&&&&&&&&&1\\ \\ \end{array}\right)},\nn\\
\hat V_{[4,1,1]}=\scriptsize{\left(\begin{array}{ccccccccccc}\\1\\&-\cfrac{1}{[3]}&\cfrac{\sqrt{[2][4]}}{[3]}\\ \\&\cfrac{\sqrt{[2][4]}}{[3]}&\cfrac{1}{[3]}\\  \\&&&1 \\&&&&-\cfrac{1}{[3]}&\cfrac{\sqrt{[2][4]}}{[3]}\\ \\&&&&\cfrac{\sqrt{[2][4]}}{[3]}&\cfrac{1}{[3]}\\ &&&&&&1\\&&&&&&&-1 \\ &&&&&&&&-\cfrac{1}{[3]}&\cfrac{\sqrt{[2][4]}}{[3]}\\ \\ &&&&&&&&\cfrac{\sqrt{[2][4]}}{[3]}&\cfrac{1}{[3]}\\ \\\end{array}\right)},\nn \\
\nn
\ee
\be
 \hat W_{[4,1,1]}=\scriptsize{\left(\begin{array}{ccccccccccc}\\-\cfrac{1}{[4]}&\cfrac{\sqrt{[3][5]}}{[4]}\\ \\ \cfrac{\sqrt{[3][5]}}{[4]}&\cfrac{1}{[4]}\\ \\&&1\\&&&1\\&&&&-1\\ &&&&&-\cfrac{1}{[4]}&&\cfrac{\sqrt{[3][5]}}{[4]}\\ &&&&&&-\cfrac{1}{[4]}&&\cfrac{\sqrt{[3][5]}}{[4]}\\
&&&&&\cfrac{\sqrt{[3][5]}}{[4]}&&\cfrac{1}{[4]}\\ &&&&&&\cfrac{\sqrt{[3][5]}}{[4]}&&\cfrac{1}{[4]}\\ \\&&&&&&&&&1\\  \\ \end{array}\right)},\nn\\
\hat Y_{[4,1,1]}=\scriptsize{\left(\begin{array}{ccccccccccc}\\-1 \\&-\cfrac{1}{[5]}&&&\cfrac{\sqrt{[4][6]}}{[5]}\\ &&-\cfrac{1}{[5]}&&&\cfrac{\sqrt{[4][6]}}{[5]}\\ &&&-\cfrac{1}{[5]}&&&\cfrac{\sqrt{[4][6]}}{[5]}\\ &\cfrac{\sqrt{[4][6]}}{[5]}&&&\cfrac{1}{[5]} \\&&\cfrac{\sqrt{[4][6]}}{[5]}&&&\cfrac{1}{[5]}\\ &&&\cfrac{\sqrt{[4][6]}}{[5]}&&&\cfrac{1}{[5]}\\ \\&&&&&&&1\\&&&&&&&&1\\&&&&&&&&&1\\ \\ \end{array}\right)}
\nn\ee

\bigskip

{\bf $16\times 16$ mixing matrices for $[3,2,1]$}
\be
\hat\hotR_{[3,2,1]}=\footnotesize{\left(\begin{array}{cccccccccccccccc}\\q\\&q\\&&-\cfrac{1}{q}\\&&&q\\&&&&-\cfrac{1}{q}\\&&&&&q\\
&&&&&&q\\&&&&&&&-\cfrac{1}{q}\\
&&&&&&&&q\\&&&&&&&&&-\cfrac{1}{q}\\&&&&&&&&&&-\cfrac{1}{q}\\&&&&&&&&&&&q\\&&&&&&&&&&&&-\cfrac{1}{q}\\&&&&&&&&&&&&&q\\
&&&&&&&&&&&&&&-\cfrac{1}{q}\\&&&&&&&&&&&&&&&-\cfrac{1}{q}\\ \\ \end{array}\right)},\ \nn\ee
\be
\hat U_{[3,2,1]}= \scriptsize{\left( \begin {array}{cccccccccccccccc}\\  1&&&&&&&&&&&&&&
&\\ \\&- \cfrac{1}{[2]}&{\cfrac {\sqrt {[3] }}{[2] }}&&&&&&&&&&&&&
\\ \\&-{\cfrac {\sqrt {[3] }}{[2] }}&- \cfrac{1}{[2]}&&&&&&&&&&&&&
\\ \\&&&- \cfrac{1}{[2]}&{\cfrac {\sqrt {[3] }}{[2] }}&&&&&&&&&&&
\\ \\&&&-\cfrac {\sqrt {[3]}}{[2]}&- \cfrac{1}{[2]}&&&&&&&&&&&
\\ \\&&&&&1&&&&&&&&&&
\\&&&&&&- \cfrac{1}{[2]}&{\cfrac {\sqrt {[3] }}{[2] }}&&&&&&&&
\\ \\&&&&&&-{\cfrac {\sqrt {[3] }}{[2] }}&-
 \cfrac{1}{[2]}&&&&&&&&
\\ \\&&&&&&&&- \cfrac{1}{[2]}&{\cfrac {\sqrt {[3] }}{[2] }}&&&&&&
\\ \\&&&&&&&&-{\cfrac {\sqrt {[3] }}{[2] }}&-
 \cfrac{1}{[2]}&&&&&&
\\ \\&&&&&&&&&&1&&&&&
\\&&&&&&&&&&&- \cfrac{1}{[2]}&{\cfrac {\sqrt {[3] }}{[2] }}&&&
\\ \\&&&&&&&&&&&-{\cfrac {\sqrt {[3] }}{[2] }}&- \cfrac{1}{[2]}&&&
\\ \\&&&&&&&&&&&&&- \cfrac{1}{[2]}&\cfrac {\sqrt {[3] }}{[2] }&
\\ \\&&&&&&&&&&&&&-{\cfrac {\sqrt {[3] }}{[2] }}&- \cfrac{1}{[2]}&
\\ \\&&&&&&&&&&&&&&&1\\ \\\end{array}
\right)},
\nn\ee
\be
\hat V_{[3,2,1]}= \scriptsize{\left( \begin{array}{cccccccccccccccc} \\- \cfrac{1}{[3]}&{\cfrac {\sqrt {[4] [2] }}{[3] }}&&&&&&&&&
&&&&&\\ \\ \cfrac {\sqrt {[4] [2] }}{[3] }&
 \cfrac{1}{[3]}&&&&&&&&&&&
&&&\\ &&1&&&&&&&&&&&&&
\\ &&&1&&&&&&&&&&&&
\\ &&&&-1&&&&&&&&&&&
\\ &&&&&- \cfrac{1}{[3]}&{\cfrac {\sqrt {[4] [2] }}{[3] }}&&&&&&&&&
\\ \\ &&&&&{\cfrac {\sqrt {[4] [2] }}{[3] }}&
 \cfrac{1}{[3]}&&&&&&&&&
\\ \\ &&&&&&&1&&&&&&&&
\\ &&&&&&&&-1&&&&&&&
\\ &&&&&&&&&- \cfrac{1}{[3]}&{\cfrac {\sqrt {[4] [2] }}{[3] }}&&&&&
\\ \\ &&&&&&&&&{\cfrac {\sqrt {[4] [2] }}{[3] }}&
 \cfrac{1}{[3]}&&&&&
\\ \\ &&&&&&&&&&&1&&&&
\\  &&&&&&&&&&&&-1&&&
\\  &&&&&&&&&&&&&-1&&
\\ &&&&&&&&&&&&&&- \cfrac{1}{[3]}&{\cfrac {\sqrt {[4] [2] }}{[3] }}
\\ \\ &&&&&&&&&&&&&&{\cfrac {\sqrt {[4] [2] }}{[3] }}& \cfrac{1}{[3]}\\ \\
\end{array} \right)},\nn
\ee
\\
\\
\\
\\
\be
W_{[3,2,1]}= \scriptsize{\left(\begin{array}{cccccccccccccccc}\\ 1&&&&&&&&&&&&&&
&\\ \\ \noalign{\medskip}&-\cfrac{1}{[2]}&&{\cfrac {\sqrt { [3] }}{[2] }}&&&&&&&&&&&&
\\ &&-\cfrac{1}{[2]}&&{\cfrac {\sqrt {[3]}}{[2] }}&&&&&&&&&&&
\\ &{\cfrac {\sqrt {[3]}}{[2] }}&& \cfrac{1}{[2]} &&&&&&&&&&&&
\\ &&{\cfrac {\sqrt {[3]
}}{[2] }}&& \cfrac{1}{[2]} &&&&&&&&&&&
\\ \\&&&&&-1&&&&&&&&&&
\\ \\&&&&&&-\cfrac{1}{[4]}&&\cfrac {\sqrt {[5][3]}}{[4] }&&&&&&&
\\ &&&&&&&-\cfrac{1}{[4]}&&\cfrac {\sqrt {[5][3]}}{[4] }&&&&&&
\\ &&&&&&\cfrac {\sqrt {[5][3]}}{[4] }&&
\cfrac{1}{[4]}&&&&&&&
\\ &&&&&&&\cfrac {\sqrt {[5][3]}}{[4] }&&
\cfrac{1}{[4]}&&&&&&
\\ \\&&&&&&&&&&1&&&&&
\\ \\&&&&&&&&&&&- \cfrac{1}{[2]} &&{\cfrac {\sqrt {[3] }}{[2] }}&&
\\ &&&&&&&&&&&&-\cfrac{1}{[2]}&&{\cfrac {\sqrt {[3]}}{[2] }}&
\\ &&&&&&&&&&&\cfrac {\sqrt {[3]}}{[2]}&&\cfrac{1}{[2]} &&
\\ &&&&&&&&&&&&\cfrac {\sqrt {[3]}}{[2]}&&\cfrac{1}{[2]} &
\\ \\ &&&&&&&&&&&&&&&-1\\ \\ \end{array}
\right)},
\nn\ee
\be
\hspace{-1cm}\hat Y_{[3,2,1]}=\scriptsize{\left(\begin{array}{cccccccccccccccc}\\ -  \cfrac{1}{[2]}&&&&&{\cfrac {\sqrt { [3]
 }}{ [2]  }}&&
&&&&&&&&\\ &-  \cfrac{1}{[2]}&&&&&{\cfrac {\sqrt { [3]
  }}{ [2]  }}&&
&&&&&&&\\ &&-  \cfrac{1}{[2]}&&&&&{\cfrac {\sqrt { [3]
}}{ [2]  }}&&
&&&&&&\\ &&&-  \cfrac{1}{[4]}&&&&&&&&{\cfrac {\sqrt { [5]
   [3]  }}{ [4]  }}&&
&&\\ &&&&-  \cfrac{1}{[4]}&&&&&&&&{\cfrac {\sqrt { [5]
 [3]  }}{ [4]  }}&&&
\\ \noalign{\medskip}{\cfrac {\sqrt { [3]
}}{ [2]  }}&&&&&  \cfrac{1}{[2]}&&&&&&&&&&
\\ &{\cfrac {\sqrt { [3]}}{ [2]  }}&&&&&  \cfrac{1}{[2]}&&&&&&&&&
\\ &&{\cfrac {\sqrt { [3]}}{ [2]  }}&&&&&
  \cfrac{1}{[2]}&&&&&&&&
\\ &&&&&&&&-  \cfrac{1}{[2]}&&&&&{\cfrac {\sqrt { [3]
   }}{ [2]  }}&&
\\ &&&&&&&&&-  \cfrac{1}{[2]}&&&&&{\cfrac {\sqrt { [3]
   }}{ [2]  }}&
\\ &&&&&&&&&&-  \cfrac{1}{[2]}&&&&&{\cfrac {\sqrt { [3]
   }}{ [2]  }}
\\ &&&{\cfrac {\sqrt { [5]   [3]  }}{ [4]  }}&&&&&&&
&  \cfrac{1}{[4]}&&&&
\\ &&&&{\cfrac {\sqrt { [5]
 [3]  }}{ [4]  }}&&&&&&&
&  \cfrac{1}{[4]}&&&
\\ &&&&&&&&{\cfrac {\sqrt { [3]}}{ [2]  }}&&
&&&  \cfrac{1}{[2]}&&
\\ &&&&&&&&&{\cfrac {\sqrt { [3]}}{ [2] }}&&
&&&  \cfrac{1}{[2]}&
\\ &&&&&&&&&&{\cfrac {\sqrt {
 [3]  }}{ [2]
  }}&&&&&  \cfrac{1}{[2]}\\ \\
\end{array}\right)}
\nn\ee

\section*{Appendix 2. Mixing matrices for the 7-strand formula
(\ref{7str})}

{\scriptsize $$\boxed{c_k\equiv\ \cfrac{1}{[k]},\ \ s_k\equiv\ \cfrac{\sqrt{[k-1][k+1]}}{[k]}}$$}
\tiny
$${\small \hat U_{61}=}\left(\begin{array}{cccccc}1& & & & & \\& 1& & & & \\& & 1& & & \\
& & & 1& & \\& & & & -c_2& -s_2\\& & & & s_2& -c_2\end{array}\right)$$

$${\small \hat V_{61}=}\left(\begin{array}{cccccc}1& & & & & \\& 1& & & & \\& & 1& & & \\
& & & -c_3& s_3& \\& & & s_3& c_3& \\& & & & & 1\end{array}\right)$$

$${\small \hat W_{61}=}\left(\begin{array}{cccccc}1& & & & & \\& 1& & & & \\& & -c_4& s_4& & \\
& & s_4& c_4& & \\& & & & 1& \\& & & & & 1\end{array}\right)$$

$${\small \hat Y_{61}=}\left(\begin{array}{cccccc}1& & & & & \\& -c_5& s_5& & & \\& s_5& c_5& & & \\
& & & 1& & \\& & & & 1& \\& & & & & 1\end{array}\right)$$

$${\small \hat Z_{61}=}\left(\begin{array}{cccccc}-c_6& s_6& & & & \\s_6& c_6& & & & \\& & 1& & & \\
& & & 1& & \\& & & & 1& \\& & & & & 1\end{array}\right)$$

$${\small \hat U_{52}=}\left(\begin{array}{cccccccccccccc}
1& & & & & & & & & & & & & \\
& 1& & & & & & & & & & & & \\
& & 1& & & & & & & & & & & \\
& & & -c_2& -s_2& & & & & & & & & \\
& & & s_2& -c_2& & & & & & & & & \\
& & & & & 1& & & & & & & & \\
& & & & & & 1& & & & & & & \\
& & & & & & & -c_2& -s_2& & & & & \\
& & & & & & & s_2& -c_2& & & & & \\
& & & & & & & & & 1& & & & \\
& & & & & & & & & & -c_2& -s_2& & \\
& & & & & & & & & & s_2& -c_2& & \\
& & & & & & & & & & & & -c_2& -s_2\\
& & & & & & & & & & & & s_2& -c_2\end{array}\right)$$

$${\small \hat V_{52}=}\left(\begin{array}{cccccccccccccc}
1& & & & & & & & & & & & & \\
& 1& & & & & & & & & & & & \\
& & -c_3& s_3& & & & & & & & & & \\
& & s_3& c_3& & & & & & & & & & \\
& & & & 1& & & & & & & & & \\
& & & & & 1& & & & & & & & \\
& & & & & & -c_3& s_3& & & & & & \\
& & & & & & s_3& c_3& & & & & & \\
& & & & & & & & 1& & & & & \\
& & & & & & & & & -c_3& s_3& & & \\
& & & & & & & & & s_3& c_3& & & \\
& & & & & & & & & & & 1& & \\
& & & & & & & & & & & & 1& \\
& & & & & & & & & & & & & -1\end{array}\right)$$

$${\small \hat W_{52}=}\left(\begin{array}{cccccccccccccc}
1& & & & & & & & & & & & & \\
& -c_4& s_4& & & & & & & & & & & \\
& s_4& c_4& & & & & & & & & & & \\
& & & 1& & & & & & & & & & \\
& & & & 1& & & & & & & & & \\
& & & & & -c_4& s_4& & & & & & & \\
& & & & & s_4& c_4& & & & & & & \\
& & & & & & & 1& & & & & & \\
& & & & & & & & 1& & & & & \\
& & & & & & & & & 1& & & & \\
& & & & & & & & & & -c_2& & s_2& \\
& & & & & & & & & & & -c_2& & s_2\\
& & & & & & & & & & s_2& & c_2& \\
& & & & & & & & & & & s_2& & c_2\end{array}\right)$$

$${\small \hat Y_{52}=}\left(\begin{array}{cccccccccccccc}
-c_5& s_5& & & & & & & & & & & & \\
s_5& c_5& & & & & & & & & & & & \\
& & 1& & & & & & & & & & & \\
& & & 1& & & & & & & & & & \\
& & & & 1& & & & & & & & & \\
& & & & & 1& & & & & & & & \\
& & & & & & -c_3& & & s_3& & & & \\
& & & & & & & -c_3& & & s_3& & & \\
& & & & & & & & -c_3& & & s_3& & \\
& & & & & & s_3& & & c_3& & & & \\
& & & & & & & s_3& & & c_3& & & \\
& & & & & & & & s_3& & & c_3& & \\
& & & & & & & & & & & & 1& \\
& & & & & & & & & & & & & 1\end{array}\right)$$

$${\small \hat Z_{52}=}\left(\begin{array}{cccccccccccccc}
1& & & & & & & & & & & & & \\
& -c_4& & & & s_4& & & & & & & & \\
& & -c_4& & & & s_4& & & & & & & \\
& & & -c_4& & & & s_4& & & & & & \\
& & & & -c_4& & & & s_4& & & & & \\
& s_4& & & & c_4& & & & & & & & \\
& & s_4& & & & c_4& & & & & & & \\
& & & s_4& & & & c_4& & & & & & \\
& & & & s_4& & & & c_4& & & & & \\
& & & & & & & & & 1& & & & \\
& & & & & & & & & & 1& & & \\
& & & & & & & & & & & 1& & \\
& & & & & & & & & & & & 1& \\
& & & & & & & & & & & & & 1\end{array}\right)$$

$${\small \hat U_{511}=}\left(\begin{array}{ccccccccccccccc}
1& & & & & & & & & & & & & & \\
& 1& & & & & & & & & & & & & \\
& & 1& & & & & & & & & & & & \\
& & & -c_2& -s_2& & & & & & & & & & \\
& & & s_2& -c_2& & & & & & & & & & \\
& & & & & 1& & & & & & & & & \\
& & & & & & 1& & & & & & & & \\
& & & & & & & -c_2& -s_2& & & & & & \\
& & & & & & & s_2& -c_2& & & & & & \\
& & & & & & & & & 1& & & & & \\
& & & & & & & & & & -c_2& -s_2& & & \\
& & & & & & & & & & s_2& -c_2& & & \\
& & & & & & & & & & & & -c_2& -s_2& \\
& & & & & & & & & & & & s_2& -c_2& \\
& & & & & & & & & & & & & & -1\end{array}\right)$$

$${\small \hat V_{511}=}\left(\begin{array}{ccccccccccccccc}
1& & & & & & & & & & & & & & \\
 & 1& & & & & & & & & & & & & \\
& & -c_3& s_3& & & & & & & & & & & \\
& & s_3& c_3& & & & & & & & & & & \\
& & & & 1& & & & & & & & & & \\
& & & & & 1& & & & & & & & & \\
& & & & & & -c_3& s_3& & & & & & & \\
& & & & & & s_3& c_3& & & & & & & \\
& & & & & & & & 1& & & & & & \\
& & & & & & & & & -c_3& s_3& & & & \\
& & & & & & & & & s_3& c_3& & & & \\
& & & & & & & & & & & 1& & & \\
& & & & & & & & & & & & -1& & \\
& & & & & & & & & & & & & -c_3& s_3\\
& & & & & & & & & & & & & s_3& c_3\end{array}\right)$$

$${\small \hat W_{511}=}\left(\begin{array}{ccccccccccccccc}
1& & & & & & & & & & & & & & \\
& -c_4& s_4& & & & & & & & & & & & \\
& s_4& c_4& & & & & & & & & & & & \\
& & & 1& & & & & & & & & & & \\
& & & & 1& & & & & & & & & & \\
& & & & & -c_4& s_4& & & & & & & & \\
& & & & & s_4& c_4& & & & & & & & \\
& & & & & & & 1& & & & & & & \\
& & & & & & & & 1& & & & & & \\
& & & & & & & & & -1& & & & & \\
& & & & & & & & & & -c_4& & s_4& & \\
& & & & & & & & & & & -c_4& & s_4& \\
& & & & & & & & & & s_4& & c_4& & \\
& & & & & & & & & & & s_4& & c_4& \\
& & & & & & & & & & & & & & 1\end{array}\right)$$

$${\small \hat Y_{511}=}\left(\begin{array}{ccccccccccccccc}
-c_5& s_5& & & & & & & & & & & & & \\
s_5& c_5& & & & & & & & & & & & & \\
& & 1& & & & & & & & & & & & \\
& & & 1& & & & & & & & & & & \\
& & & & 1& & & & & & & & & & \\
& & & & & -1& & & & & & & & & \\
& & & & & & -c_5& & & s_5& & & & & \\
& & & & & & & -c_5& & & s_5& & & & \\
& & & & & & & & -c_5& & & s_5& & & \\
& & & & & & s_5& & & c_5& & & & & \\
& & & & & & & s_5& & & c_5& & & & \\
& & & & & & & & s_5& & & c_5& & & \\
& & & & & & & & & & & & 1& & \\
& & & & & & & & & & & & & 1& \\
& & & & & & & & & & & & & & 1\end{array}\right)$$

$${\small \hat Z_{511}=}\left(\begin{array}{ccccccccccccccc}
-1& & & & & & & & & & & & & & \\
& -c_6& & & & s_6& & & & & & & & & \\
& & -c_6& & & & s_6& & & & & & & & \\
& & & -c_6& & & & s_6& & & & & & & \\
& & & & -c_6& & & & s_6& & & & & & \\
& s_6& & & & c_6& & & & & & & & & \\
& & s_6& & & & c_6& & & & & & & & \\
& & & s_6& & & & c_6& & & & & & & \\
& & & & s_6& & & & c_6& & & & & & \\
& & & & & & & & & 1& & & & & \\
& & & & & & & & & & 1& & & & \\
& & & & & & & & & & & 1& & & \\
& & & & & & & & & & & & 1& & \\
& & & & & & & & & & & & & 1& \\
& & & & & & & & & & & & & & 1\end{array}\right)$$
\newpage
$${\small \hat U_{43}=}\left(\begin{array}{ccccccccccccccc}
1& & & & & & & & & & & & & \\
& 1& & & & & & & & & & & & \\
& & -c_2& -s_2& & & & & & & & & & \\
& & s_2& -c_2& & & & & & & & & & \\
& & & & 1& & & & & & & & & \\
& & & & & -c_2& -s_2& & & & & & & \\
& & & & & s_2& -c_2& & & & & & & \\
& & & & & & & -c_2& -s_2& & & & & \\
& & & & & & & s_2& -c_2& & & & & \\
& & & & & & & & & 1& & & & \\
& & & & & & & & & & -c_2& -s_2& & \\
& & & & & & & & & & s_2& -c_2& & \\
& & & & & & & & & & & & -c_2& -s_2\\
& & & & & & & & & & & & s_2& -c_2\end{array}\right)$$

$${\small \hat V_{43}=}\left(\begin{array}{cccccccccccccc}
1& & & & & & & & & & & & & \\
& -c_3& s_3& & & & & & & & & & & \\
& s_3& c_3& & & & & & & & & & & \\
& & & 1& & & & & & & & & & \\
& & & & -c_3& s_3& & & & & & & & \\
& & & & s_3& c_3& & & & & & & & \\
& & & & & & 1& & & & & & & \\
& & & & & & & 1& & & & & & \\
& & & & & & & & -1& & & & & \\
& & & & & & & & & -c_3& s_3& & & \\
& & & & & & & & & s_3& c_3& & & \\
& & & & & & & & & & & 1& & \\
& & & & & & & & & & & & 1& \\
& & & & & & & & & & & & & -1\end{array}\right)$$

$${\small \hat W_{43}=}\left(\begin{array}{cccccccccccccc}
-c_4& s_4& & & & & & & & & & & & \\
s_4& c_4& & & & & & & & & & & & \\
& & 1& & & & & & & & & & & \\
& & & 1& & & & & & & & & & \\
& & & & 1& & & & & & & & & \\
& & & & & -c_2& & s_2& & & & & & \\
& & & & & & -c_2& & s_2& & & & & \\
& & & & & s_2& & c_2& & & & & & \\
& & & & & & s_2& & c_2& & & & & \\
& & & & & & & & & 1& & & & \\
& & & & & & & & & & -c_2& & s_2& \\
& & & & & & & & & & & -c_2& & s_2\\
& & & & & & & & & & s_2& & c_2& \\
& & & & & & & & & & & s_2& & c_2\end{array}\right)$$

$${\small \hat Y_{43}=}\left(\begin{array}{cccccccccccccc}
1& & & & & & & & & & & & & \\
& -c_3& & & s_3& & & & & & & & & \\
& & -c_3& & & s_3& & & & & & & & \\
& & & -c_3& & & s_3& & & & & & & \\
& s_3& & & c_3& & & & & & & & & \\
& & s_3& & & c_3& & & & & & & & \\
& & & s_3& & & c_3& & & & & & & \\
& & & & & & & 1& & & & & & \\
& & & & & & & & 1& & & & & \\
& & & & & & & & & 1& & & & \\
& & & & & & & & & & 1& & & \\
& & & & & & & & & & & 1& & \\
& & & & & & & & & & & & -1& \\
& & & & & & & & & & & & & -1\end{array}\right)$$

$${\small \hat Z_{43}=}\left(\begin{array}{cccccccccccccc}
1& & & & & & & & & & & & & \\
& 1& & & & & & & & & & & & \\
& & 1& & & & & & & & & & & \\
& & & 1& & & & & & & & & & \\
& & & & -c_2& & & & & s_2& & & & \\
& & & & & -c_2& & & & & s_2& & & \\
& & & & & & -c_2& & & & & s_2& & \\
& & & & & & & -c_2& & & & & s_2& \\
& & & & & & & & -c_2& & & & & s_2\\
& & & & s_2& & & & & c_2& & & & \\
& & & & & s_2& & & & & c_2& & & \\
& & & & & & s_2& & & & & c_2& & \\
& & & & & & & s_2& & & & & c_2& \\
& & & & & & & & s_2& & & & & c_2\end{array}\right)$$

\rotate{${\small \hat U_{421}=}$ $\left(\begin{array}{ccccccccccccccccccccccccccccccccccc}
1& & & & & & & & & & & & & & & & & & & & & & & & &
& & & & & & & & & \\& 1& & & & & & & & & & & & & &
& & & & & & & & & & & & & & & & & & & \\& & -c_2&
-s_2& & & & & & & & & & & & & & & & & & & & & & & &
& & & & & & & \\& & s_2& -c_2& & & & & & & & & & &
& & & & & & & & & & & & & & & & & & & & \\& & & &
1& & & & & & & & & & & & & & & & & & & & & & & & &
& & & & & \\& & & & & -c_2& -s_2& & & & & & & & & &
& & & & & & & & & & & & & & & & & & \\& & & & &
s_2& -c_2& & & & & & & & & & & & & & & & & & & & & &
& & & & & & \\& & & & & & & -c_2& -s_2& & & & & & &
& & & & & & & & & & & & & & & & & & & \\
\end{array}\right)$}

\rotate{${\small \hat V_{421}=}$ $\left(\begin{array}{ccccccccccccccccccccccccccccccccccc}
1& & & & & & & & & & & & & & & & & & & & & & & & &
& & & & & & & & & \\& -c_3& s_3& & & & & & & & & &
& & & & & & & & & & & & & & & & & & & & & & \\&
s_3& c_3& & & & & & & & & & & & & & & & & & & & & &
& & & & & & & & & & \\& & & 1& & & & & & & & & & &
& & & & & & & & & & & & & & & & & & & & \\& & & &
-c_3& s_3& & & & & & & & & & & & & & & & & & & & & &
& & & & & & & \\& & & & s_3& c_3& & & & & & & & & &
& & & & & & & & & & & & & & & & & & & \\& & & & &
& 1& & & & & & & & & & & & & & & & & & & & & & & &
& & & & \\& & & & & & & 1& & & & & & & & & & & & &
& & & & & & & & & & & & & & \\
\end{array}\right)$}

\rotate{${\small \hat W_{421}=}$ $\left(\begin{array}{ccccccccccccccccccccccccccccccccccc}
-c_4& s_4& & & & & & & & & & & & & & & & & & & & & &
& & & & & & & & & & & \\s_4& c_4& & & & & & & & & &
& & & & & & & & & & & & & & & & & & & & & & & \\&
& 1& & & & & & & & & & & & & & & & & & & & & & & &
& & & & & & & & \\& & & 1& & & & & & & & & & & & &
& & & & & & & & & & & & & & & & & & \\& & & & 1& &
& & & & & & & & & & & & & & & & & & & & & & & & &
& & & \\& & & & & -c_2& & s_2& & & & & & & & & & &
& & & & & & & & & & & & & & & & \\& & & & & &
-c_2& & s_2& & & & & & & & & & & & & & & & & & & & &
& & & & & \\& & & & & s_2& & c_2& & & & & & & & & &
& & & & & & & & & & & & & & & & & \\
\end{array}\right)$}

\rotate{${\small \hat Y_{421}=}$ $\left(\begin{array}{ccccccccccccccccccccccccccccccccccc}
1& & & & & & & & & & & & & & & & & & & & & & & & &
& & & & & & & & & \\& -c_3& & & s_3& & & & & & & &
& & & & & & & & & & & & & & & & & & & & & & \\& &
-c_3& & &s_3& & & & & & & & & & & & & & & & & & & &
& & & & & & & & & \\& & & -c_3& & & s_3& & & & & &
& & & & & & & & & & & & & & & & & & & & & & \\&
s_3& & & c_3& & & & & & & & & & & & & & & & & & & &
& & & & & & & & & & \\& & s_3& & & c_3& & & & & & &
& & & & & & & & & & & & & & & & & & & & & & \\& &
& s_3& & & c_3& & & & & & & & & & & & & & & & & & &
& & & & & & & & & \\& & & & & & & 1& & & & & & & &
& & & & & & & & & & & & & & & & & & & \\
\end{array}\right)$}

\rotate{{$\small \hat Z_{421}=$}$\left(\begin{array}{ccccccccccccccccccccccccccccccccccc}
-c_2& & & & & & & & & s_2& & & & & & & & & & & & & &
& & & & & & & & & & & \\& -c_2& & & & & & & & &
s_2& & & & & & & & & & & & & & & & & & & & & & & &
\\& & -c_2& & & & & & & & & s_2& & & & & & & & & & &
& & & & & & & & & & & & \\& & & -c_2& & & & & & &
& & s_2& & & & & & & & & & & & & & & & & & & & & &
\\& & & & -c_5& & & & & & & & & & & & & & & s_5& & &
& & & & & & & & & & & & \\& & & & & -c_5& & & & &
& & & & & & & & & & s_5& & & & & & & & & & & & & &
\\& & & & & & -c_5& & & & & & & & & & & & & & & s_5&
& & & & & & & & & & & & \\& & & & & & & -c_5& & &
& & & & & & & & & & & & s_5& & & & & & & & & & & &
\\
\end{array}\right)$}

$${\small \hat U_{4111}=}\left(\begin{array}{cccccccccccccccccccc}
1& & & & & & & & & & & & & & & & & & & \\
& 1& & & & & & & & & & & & & & & & & & \\
& & -c_2& -s_2& & & & & & & & & & & & & & & & \\
& & s_2& -c_2& & & & & & & & & & & & & & & & \\
& & & & 1& & & & & & & & & & & & & & & \\
& & & & & -c_2& -s_2& & & & & & & & & & & & & \\
& & & & & s_2& -c_2& & & & & & & & & & & & & \\
& & & & & & & -c_2& -s_2& & & & & & & & & & & \\
& & & & & & & s_2& -c_2& & & & & & & & & & & \\
& & & & & & & & & -1& & & & & & & & & & \\
& & & & & & & & & & 1& & & & & & & & & \\
& & & & & & & & & & & -c_2& -s_2& & & & & & & \\
& & & & & & & & & & & s_2& -c_2& & & & & & & \\
& & & & & & & & & & & & & -c_2& -s_2& & & & & \\
& & & & & & & & & & & & & s_2& -c_2& & & & & \\
& & & & & & & & & & & & & & & -1& & & & \\
& & & & & & & & & & & & & & & & -c_2& -s_2& & \\
& & & & & & & & & & & & & & & & s_2& -c_2& & \\
& & & & & & & & & & & & & & & & & & -1& \\
& & & & & & & & & & & & & & & & & & & -1\end{array}\right)$$

$${\small \hat V_{4111}=}\left(\begin{array}{cccccccccccccccccccc}
1& & & & & & & & & & & & & & & & & & & \\
& -c_3& s_3& & & & & & & & & & & & & & & & & \\
& s_3& c_3& & & & & & & & & & & & & & & & & \\
& & & 1& & & & & & & & & & & & & & & & \\
& & & & -c_3& s_3& & & & & & & & & & & & & & \\
& & & & s_3& c_3& & & & & & & & & & & & & & \\
& & & & & & 1& & & & & & & & & & & & & \\
& & & & & & & -1& & & & & & & & & & & & \\
& & & & & & & & -c_3& s_3& & & & & & & & & & \\
& & & & & & & & s_3& c_3& & & & & & & & & & \\
& & & & & & & & & & -c_3& s_3& & & & & & & & \\
& & & & & & & & & & s_3& c_3& & & & & & & & \\
& & & & & & & & & & & & 1& & & & & & & \\
& & & & & & & & & & & & & -1& & & & & & \\
& & & & & & & & & & & & & & -c_3& s_3& & & & \\
& & & & & & & & & & & & & & s_3& c_3& & & & \\
& & & & & & & & & & & & & & & & -1& & & \\
& & & & & & & & & & & & & & & & & -c_3& s_3& \\
& & & & & & & & & & & & & & & & & s_3& c_3& \\
& & & & & & & & & & & & & & & & & & & -1\end{array}\right)$$

$${\small \hat W_{4111}=}\left(\begin{array}{cccccccccccccccccccc}
-c_4& s_4& & & & & & & & & & & & & & & & & & \\
s_4& c_4& & & & & & & & & & & & & & & & & & \\
& & 1& & & & & & & & & & & & & & & & & \\
& & & 1& & & & & & & & & & & & & & & & \\
& & & & -1& & & & & & & & & & & & & & & \\
& & & & & -c_4& & s_4& & & & & & & & & & & & \\
& & & & & & -c_4& & s_4& & & & & & & & & & & \\
& & & & & s_4& & c_4& & & & & & & & & & & & \\
& & & & & & s_4& & c_4& & & & & & & & & & & \\
& & & & & & & & & 1& & & & & & & & & & \\
& & & & & & & & & & -1& & & & & & & & & \\
& & & & & & & & & & & -c_4& & s_4& & & & & & \\
& & & & & & & & & & & & -c_4& & s_4& & & & & \\
& & & & & & & & & & & s_4& & c_4& & & & & & \\
& & & & & & & & & & & & s_4& & c_4& & & & & \\
& & & & & & & & & & & & & & & 1& & & & \\
& & & & & & & & & & & & & & & & -1& & & \\
& & & & & & & & & & & & & & & & & -1& & \\
& & & & & & & & & & & & & & & & & & -c_4& s_4\\
& & & & & & & & & & & & & & & & & & s_4& c_4\end{array}\right)$$

$${\small \hat Y_{4111}=}\left(\begin{array}{cccccccccccccccccccc}
-1& & & & & & & & & & & & & & & & & & & \\
& -c_5& & & s_5& & & & & & & & & & & & & & & \\
& & -c_5& & & s_5& & & & & & & & & & & & & & \\
& & & -c_5& & & s_5& & & & & & & & & & & & & \\
& s_5& & & c_5& & & & & & & & & & & & & & & \\
& & s_5& & & c_5& & & & & & & & & & & & & & \\
& & & s_5& & & c_5& & & & & & & & & & & & & \\
& & & & & & & 1& & & & & & & & & & & & \\
& & & & & & & & 1& & & & & & & & & & & \\
& & & & & & & & & 1& & & & & & & & & & \\
& & & & & & & & & & -1& & & & & & & & & \\
& & & & & & & & & & & -1& & & & & & & & \\
& & & & & & & & & & & & -1& & & & & & & \\
& & & & & & & & & & & & & -c_5& & & s_5& & & \\
& & & & & & & & & & & & & & -c_5& & & s_5& & \\
& & & & & & & & & & & & & & & -c_5& & & s_5& \\
& & & & & & & & & & & & & s_5& & & c_5& & & \\
& & & & & & & & & & & & & & s_5& & & c_5& & \\
& & & & & & & & & & & & & & & s_5& & & c_5& \\
 & & & & & & & & & & & & & & & & & & & 1\end{array}\right)$$

$${\small \hat Z_{4111}=}\left(\begin{array}{cccccccccccccccccccc}
-1& & & & & & & & & & & & & & & & & & & \\
& -1& & & & & & & & & & & & & & & & & & \\
& & -1& & & & & & & & & & & & & & & & & \\
& & & -1& & & & & & & & & & & & & & & & \\
& & & & -c_6& & & & & & s_6& & & & & & & & & \\
& & & & & -c_6& & & & & & s_6& & & & & & & & \\
& & & & & & -c_6& & & & & & s_6& & & & & & & \\
& & & & & & & -c_6& & & & & & s_6& & & & & & \\
& & & & & & & & -c_6& & & & & & s_6& & & & & \\
& & & & & & & & & -c_6& & & & & & s_6& & & & \\
& & & & s_6& & & & & & c_6& & & & & & & & & \\
& & & & & s_6& & & & & & c_6& & & & & & & & \\
& & & & & & s_6& & & & & & c_6& & & & & & & \\
& & & & & & & s_6& & & & & & c_6& & & & & & \\
& & & & & & & & s_6& & & & & & c_6& & & & & \\
& & & & & & & & & s_6& & & & & & c_6& & & & \\
& & & & & & & & & & & & & & & & 1& & & \\
& & & & & & & & & & & & & & & & & 1& & \\
& & & & & & & & & & & & & & & & & & 1& \\
& & & & & & & & & & & & & & & & & & & 1\end{array}\right)$$

 $${\small \hat U_{331}=}\left(\begin{array}{ccccccccccccccccccccc}
 1& & & & & & & & & & & & & & & & & & & & \\
& -c_2& -s_2& & & & & & & & & & & & & & & & & & \\
& s_2& -c_2& & & & & & & & & & & & & & & & & & \\
& & & -c_2& -s_2& & & & & & & & & & & & & & & & \\
& & & s_2& -c_2& & & & & & & & & & & & & & & & \\
& & & & & 1& & & & & & & & & & & & & & & \\
& & & & & & -c_2& -s_2& & & & & & & & & & & & & \\
& & & & & & s_2& -c_2& & & & & & & & & & & & & \\
& & & & & & & & -c_2& -s_2& & & & & & & & & & & \\
& & & & & & & & s_2& -c_2& & & & & & & & & & & \\
& & & & & & & & & & 1& & & & & & & & & & \\
& & & & & & & & & & & -c_2& -s_2& & & & & & & & \\
& & & & & & & & & & & s_2& -c_2& & & & & & & & \\
& & & & & & & & & & & & & -c_2& -s_2& & & & & & \\
& & & & & & & & & & & & & s_2& -c_2& & & & & & \\
& & & & & & & & & & & & & & & -1& & & & & \\
& & & & & & & & & & & & & & & & -c_2& -s_2& & & \\
& & & & & & & & & & & & & & & & s_2& -c_2& & & \\
& & & & & & & & & & & & & & & & & & -c_2& -s_2& \\
& & & & & & & & & & & & & & & & & & s_2& -c_2& \\
& & & & & & & & & & & & & & & & & & & & -1\end{array}\right)$$

$${\small \hat V_{331}=}\left(\begin{array}{ccccccccccccccccccccc}
-c_3& s_3& & & & & & & & & & & & & & & & & & & \\
s_3& c_3& & & & & & & & & & & & & & & & & & & \\
& & 1& & & & & & & & & & & & & & & & & & \\
& & & 1& & & & & & & & & & & & & & & & & \\
& & & & -1& & & & & & & & & & & & & & & & \\
& & & & & -c_3& s_3& & & & & & & & & & & & & & \\
& & & & & s_3& c_3& & & & & & & & & & & & & & \\
& & & & & & & 1& & & & & & & & & & & & & \\
& & & & & & & & 1& & & & & & & & & & & & \\
& & & & & & & & & -1& & & & & & & & & & & \\
& & & & & & & & & & -c_3& s_3& & & & & & & & & \\
& & & & & & & & & & s_3& c_3& & & & & & & & & \\
& & & & & & & & & & & & 1& & & & & & & & \\
& & & & & & & & & & & & & -1& & & & & & & \\
& & & & & & & & & & & & & & -c_3& s_3& & & & & \\
& & & & & & & & & & & & & & s_3& c_3& & & & & \\
& & & & & & & & & & & & & & & & 1& & & & \\
& & & & & & & & & & & & & & & & & -1& & & \\
& & & & & & & & & & & & & & & & & & -1& & \\
& & & & & & & & & & & & & & & & & & & -c_3& s_3\\
& & & & & & & & & & & & & & & & & & & s_3& c_3\end{array}\right)$$

$${\small \hat W_{331}=}\left(\begin{array}{ccccccccccccccccccccc}
1& & & & & & & & & & & & & & & & & & & & \\
& -c_2& & s_2& & & & & & & & & & & & & & & & & \\
& & -c_2& & s_2& & & & & & & & & & & & & & & & \\
& s_2& & c_2& & & & & & & & & & & & & & & & & \\
& & s_2& & c_2& & & & & & & & & & & & & & & & \\
& & & & & 1& & & & & & & & & & & & & & & \\
& & & & & & -c_2& & s_2& & & & & & & & & & & & \\
& & & & & & & -c_2& & s_2& & & & & & & & & & & \\
& & & & & & s_2& & c_2& & & & & & & & & & & & \\
& & & & & & & s_2& & c_2& & & & & & & & & & & \\
& & & & & & & & & & -1& & & & & & & & & & \\
& & & & & & & & & & & -c_4& & s_4& & & & & & & \\
& & & & & & & & & & & & -c_4& & s_4& & & & & & \\
& & & & & & & & & & & s_4& & c_4& & & & & & & \\
& & & & & & & & & & & & s_4& & c_4& & & & & & \\
& & & & & & & & & & & & & & & 1& & & & & \\
& & & & & & & & & & & & & & & & -c_2& & s_2& & \\
& & & & & & & & & & & & & & & & & -c_2& & s_2& \\
& & & & & & & & & & & & & & & & s_2& & c_2& & \\
& & & & & & & & & & & & & & & & & s_2& & c_2& \\
& & & & & & & & & & & & & & & & & & & & -1\end{array}\right)$$

$${\small \hat Y_{331}=}\left(\begin{array}{ccccccccccccccccccccc}
1& & & & & & & & & & & & & & & & & & & & \\
& 1& & & & & & & & & & & & & & & & & & & \\
& & 1& & & & & & & & & & & & & & & & & & \\
& & & -1& & & & & & & & & & & & & & & & & \\
& & & & -1& & & & & & & & & & & & & & & & \\
& & & & & -c_2& & & & & s_2& & & & & & & & & & \\
& & & & & & -c_2& & & & & s_2& & & & & & & & & \\
& & & & & & & -c_2& & & & & s_2& & & & & & & & \\
& & & & & & & & -c_4& & & & & & & & s_4& & & & \\
& & & & & & & & & -c_4& & & & & & & & s_4& & & \\
& & & & & s_2& & & & & c_2& & & & & & & & & & \\
& & & & & & s_2& & & & & c_2& & & & & & & & & \\
& & & & & & & s_2& & & & & c_2& & & & & & & & \\
& & & & & & & & & & & & & -c_2& & & & & s_2& & \\
& & & & & & & & & & & & & & -c_2& & & & & s_2& \\
& & & & & & & & & & & & & & & -c_2& & & & & s_2\\
& & & & & & & & s_4& & & & & & & & c_4& & & & \\
& & & & & & & & & s_4& & & & & & & & c_4& & & \\
& & & & & & & & & & & & & s_2& & & & & c_2& & \\
& & & & & & & & & & & & & & s_2& & & & & c_2& \\
& & & & & & & & & & & & & & & s_2& & & & & c_2\end{array}\right)$$

$${\small \hat Z_{331}=}\left(\begin{array}{ccccccccccccccccccccc}
-c_3& & & & & s_3& & & & & & & & & & & & & & & \\
& -c_3& & & & & s_3& & & & & & & & & & & & & & \\
& & -c_3& & & & & s_3& & & & & & & & & & & & & \\
& & & -c_3& & & & & s_3& & & & & & & & & & & & \\
& & & & -c_3& & & & & s_3& & & & & & & & & & & \\
s_3& & & & & c_3& & & & & & & & & & & & & & & \\
& s_3& & & & & c_3& & & & & & & & & & & & & & \\
& & s_3& & & & & c_3& & & & & & & & & & & & & \\
& & & s_3& & & & & c_3& & & & & & & & & & & & \\
& & & & s_3& & & & & c_3& & & & & & & & & & & \\
& & & & & & & & & & 1& & & & & & & & & & \\
& & & & & & & & & & & 1& & & & & & & & & \\
& & & & & & & & & & & & 1& & & & & & & & \\
& & & & & & & & & & & & & 1& & & & & & & \\
& & & & & & & & & & & & & & 1& & & & & & \\
& & & & & & & & & & & & & & & 1& & & & & \\
& & & & & & & & & & & & & & & & -1& & & & \\
& & & & & & & & & & & & & & & & & -1& & & \\
& & & & & & & & & & & & & & & & & & -1& & \\
& & & & & & & & & & & & & & & & & & & -1& \\
& & & & & & & & & & & & & & & & & & & & -1\end{array}\right)$$

\end{document}